\documentclass[reprint, amsmath, amssymb, aps, floatfix, pra, superscriptaddress, 12pt]{revtex4-1}

\makeatletter 
    
\renewcommand\onecolumngrid{% <<<<<<
\do@columngrid{one}{\@ne}%
\def\set@footnotewidth{\onecolumngrid}% <<<<<<<<<<<<<<<<
\def\footnoterule{\kern-6pt\hrule width 1.5in\kern6pt}%
}

\renewcommand\twocolumngrid{% <<<<<<
        \def\footnoterule{% restore rule
        \dimen@\skip\footins\divide\dimen@\thr@@
        \kern-\dimen@\hrule width.5in\kern\dimen@}
        \do@columngrid{mlt}{\tw@}
}%

\def\@hangfrom@section#1#2#3{\@hangfrom{#1#2}#3}%\MakeTextUppercase{#3}}%
\def\@hangfroms@section#1#2{#1#2}%\MakeTextUppercase{#2}}%

\makeatother

\usepackage{etoolbox}
\patchcmd{\section}
  {\centering}
  {\raggedright}
  {}
  {}

\usepackage[english]{babel}
\usepackage[T1]{fontenc}
\usepackage{braket}

\usepackage{amsmath}
\usepackage{amsfonts}
\usepackage{graphicx}
\usepackage{dcolumn}
\usepackage{bm}
\usepackage[colorinlistoftodos]{todonotes}
\usepackage[colorlinks=true, allcolors=blue]{hyperref}
\usepackage[margin=0.7in]{geometry}
\usepackage{ragged2e}

\begin{document}
%\title{Planar thermal transport mapping of an epitaxial gallium nitride film}
\title{Spatiotemporal Mapping of Anisotropic Thermal Transport in GaN Thin Films via Ultrafast X-ray Diffraction}

\affiliation{Department of Nuclear Science and Engineering, Massachusetts Institute of Technology, Cambridge, MA 02139, USA}
\affiliation{Center for Computational Science and Engineering, Massachusetts Institute of Technology, Cambridge, MA 02139, USA}
\affiliation{Department of Materials Science and Engineering, Massachusetts Institute of Technology, Cambridge, MA 02139, USA}
\affiliation{Department of Mechanical Engineering, Massachusetts Institute of Technology, Cambridge, MA 02139, USA}
\affiliation{Department of Physics, Massachusetts Institute of Technology, Cambridge, MA 02139, USA}
\affiliation{Department of Electrical Engineering and Computer Science, Massachusetts Institute of Technology, Cambridge, MA 02139, USA}
\affiliation{Linac Coherent Light Source, SLAC National Accelerator Laboratory, Menlo Park, CA 94025, USA}
\affiliation{Stanford Institute for Materials and Energy Sciences, Stanford University, Stanford, CA 94025, USA}
\affiliation{Materials Science Division, Argonne National Laboratory, Lemont, IL 60439, USA}
\affiliation{MIT Media Lab, Massachusetts Institute of Technology, Cambridge, MA 02139, USA}
\affiliation{Advanced Photon Source, Argonne National Laboratory, Lemont, IL 60439, USA}
\affiliation{These authors contributed equally to this work.}

\author{Thanh Nguyen}
\thanks{Corresponding authors. \href{mailto:ngutt@mit.edu}{ngutt@mit.edu}, \href{mailto:jeehwan@mit.edu}{jeehwan@mit.edu}, \href{mailto:mingda@mit.edu}{mingda@mit.edu}}
\affiliation{Department of Nuclear Science and Engineering, Massachusetts Institute of Technology, Cambridge, MA 02139, USA}
\affiliation{These authors contributed equally to this work.}
\author{Chuliang Fu}
\affiliation{Department of Nuclear Science and Engineering, Massachusetts Institute of Technology, Cambridge, MA 02139, USA}
\affiliation{These authors contributed equally to this work.}
\author{Mouyang Cheng}
\affiliation{Center for Computational Science and Engineering, Massachusetts Institute of Technology, Cambridge, MA 02139, USA}
\affiliation{Department of Materials Science and Engineering, Massachusetts Institute of Technology, Cambridge, MA 02139, USA}
\author{Buxuan Li}
\affiliation{Department of Mechanical Engineering, Massachusetts Institute of Technology, Cambridge, MA 02139, USA}
\author{Tyra E. Espedal}
\affiliation{Department of Physics, Massachusetts Institute of Technology, Cambridge, MA 02139, USA}
\affiliation{Department of Electrical Engineering and Computer Science, Massachusetts Institute of Technology, Cambridge, MA 02139, USA}
\author{Zhantao Chen}
\affiliation{Linac Coherent Light Source, SLAC National Accelerator Laboratory, Menlo Park, CA 94025, USA}
\affiliation{Stanford Institute for Materials and Energy Sciences, Stanford University, Stanford, CA 94025, USA}
\author{Kuan Qiao}
\affiliation{Department of Mechanical Engineering, Massachusetts Institute of Technology, Cambridge, MA 02139, USA}
\author{Kumar Neeraj}
\affiliation{Materials Science Division, Argonne National Laboratory, Lemont, IL 60439, USA}
\author{Abhijatmedhi Chotrattanapituk}
\affiliation{Department of Electrical Engineering and Computer Science, Massachusetts Institute of Technology, Cambridge, MA 02139, USA}
\author{Denisse Cordova Carrizales}
\affiliation{Department of Nuclear Science and Engineering, Massachusetts Institute of Technology, Cambridge, MA 02139, USA}
\author{Eunbi Rha}
\affiliation{Department of Nuclear Science and Engineering, Massachusetts Institute of Technology, Cambridge, MA 02139, USA}
\author{Tongtong Liu}
\affiliation{Department of Physics, Massachusetts Institute of Technology, Cambridge, MA 02139, USA}
\author{Shivam N. Kajale}
\affiliation{MIT Media Lab, Massachusetts Institute of Technology, Cambridge, MA 02139, USA}
\author{Deblina Sarkar}
\affiliation{MIT Media Lab, Massachusetts Institute of Technology, Cambridge, MA 02139, USA}
\author{Donald A. Walko}
\affiliation{Advanced Photon Source, Argonne National Laboratory, Lemont, IL 60439, USA}
\author{Haidan Wen}
\affiliation{Materials Science Division, Argonne National Laboratory, Lemont, IL 60439, USA}
\affiliation{Advanced Photon Source, Argonne National Laboratory, Lemont, IL 60439, USA}
\author{Svetlana V. Boriskina}
\affiliation{Department of Mechanical Engineering, Massachusetts Institute of Technology, Cambridge, MA 02139, USA}
\author{Gang Chen}
\affiliation{Department of Mechanical Engineering, Massachusetts Institute of Technology, Cambridge, MA 02139, USA}
\author{Jeehwan Kim}
\thanks{Corresponding authors. \href{mailto:ngutt@mit.edu}{ngutt@mit.edu}, \href{mailto:jeehwan@mit.edu}{jeehwan@mit.edu}, \href{mailto:mingda@mit.edu}{mingda@mit.edu}}
\affiliation{Department of Mechanical Engineering, Massachusetts Institute of Technology, Cambridge, MA 02139, USA}
\author{Mingda Li}
\thanks{Corresponding authors. \href{mailto:ngutt@mit.edu}{ngutt@mit.edu}, \href{mailto:jeehwan@mit.edu}{jeehwan@mit.edu}, \href{mailto:mingda@mit.edu}{mingda@mit.edu}}
\affiliation{Department of Nuclear Science and Engineering, Massachusetts Institute of Technology, Cambridge, MA 02139, USA}

\date{\today}

\maketitle

\onecolumngrid

\clearpage

\noindent \textbf{Abstract}

Efficient thermal management is essential for the reliability of modern power electronics, where increasing device density leads to severe heat dissipation challenges. However, in thin-film systems, thermal transport is often compromised by interfacial resistance and microscale defects introduced during synthesis or transfer, which are difficult to characterize using conventional techniques. Here we present a non-contact, spatiotemporal-resolved ultrafast x-ray diffraction method to extract in-plane thermal conductivity and thermal boundary conductance, using GaN thin films on silicon as a model system. By tracking the pump-induced lattice strain, we reconstruct the lateral heat flow dynamics and quantitatively probe thermal transport near a wrinkle defect. We uncover pronounced asymmetric heat dissipation across the wrinkle, with a four-fold reduction in the local thermal conductivity near the wrinkle and a 25\% drop in interfacial conductance. Our work demonstrates that ultrafast x-ray diffraction can serve as a precise thermal metrology tool for characterizing heat transport in multilayered thin-film structures for next-generation microelectronic devices.\\

\noindent \textbf{Introduction}

Thermal management is crucial for the continued scaling and reliability of high-performance electronics, where the increased power densities exacerbate heat dissipation challenges \cite{sabry2019}. As devices shrink to nanoscale dimensions, accurate modeling of heat flow, particularly within a thin film, across interfaces, and in the presence of local hotspots, becomes critical to avoid thermal bottlenecks and failure modes \cite{pop2010,cahill2014}. To meet this need, a range of advanced thermometry techniques have been developed over the past few decades, including time (or frequency) domain thermoreflectance (TDTR/FDTR) \cite{cahill2004,schmidt2010,jiang2018}, the 3$\omega$ method \cite{cahill1990}, Raman thermometry \cite{malekpour2018}, scanning thermal probes \cite{nguyen2024}, and micro-fabricated thermal bridges \cite{shi2003}, among others. However, many of these techniques are constrained by factors such as surface sensitivity, the need for metal transducers, contact resistance, lack of layer-resolved information, and limited access to in-plane thermal transport, particularly in a multi-layered setting. Extensions such as beam offset transient thermoreflectance \cite{rodin2017} and steady-state thermoreflectance \cite{braun2019} techniques have improved lateral sensitivity, but challenges remain in quantifying thermal transport in buried layers. On the other hand, ultrafast electron diffraction \cite{chen2023} and ultrafast x-ray diffraction have emerged as powerful tools for probing non-equilibrium phonon dynamics \cite{thomsen1986,highland2007ballistic,zhu2017structural,nyby2020}, with demonstrated sensitivity to coherent phonon propagation \cite{sokolowski-tinten2003,lee2005}, photoinduced stress responses \cite{wen2013,schick2014}, and transient structural responses \cite{zhu2015,pudell2018}. In particular, the use of x-rays offers advantages such as deep penetration, elemental and structural specificity, and sub-picometer sensitivity to extremely small lattice constant changes, which may offer new insights on thermal metrology that are otherwise inaccessible to optical probes. 

These capabilities are especially relevant for materials like GaN, which has been widely explored for thermal management in power electronics due to its wide bandgap, high intrinsic thermal conductivity, and robust breakdown field \cite{xu2019,hoque2021,perez2023}. Thin-film GaN remotely transferred onto silicon offers a promising platform for scalable, cost-effective device integration \cite{kim2022remote}. However, the thermal performance of such heterostructures is often degraded by defects from synthesis (e.g., grain boundaries, dislocations \cite{serov2013,sood2018,li2020}) or transfer (e.g., wrinkles, etchant residues \cite{wang2014}), which are difficult to resolve using conventional techniques.

In this work, we use spatially and temporally resolved ultrafast x-ray diffraction to measure thermal transport in a GaN thin film grown by remote epitaxy onto a silicon substrate -- a platform relevant for high electron mobility transistors (HEMTs) and flexible electronic applications. By scanning the micrometer-scale spatial offset between the optical pump and x-ray probe, we track the evolution of photoinduced lattice strain, initially driven by carrier excitation and followed by a diffusive lattice thermal relaxation. Through simple thermal transport model of the long-time lattice dynamics, we extract the in-plane thermal conductivity of the GaN thin film and the thermal boundary conductance (TBC) at the GaN-Si interface, both consistent with literature and corroborated with our FDTR measurements. Importantly, the spatial resolution of the technique allows us to directly quantify heat transport across a single wrinkle defect, revealing a four-fold reduction in local thermal conductivity a 25\% drop in TBC, and a pronounced asymmetric thermal dissipation. This scanning ultrafast x-ray diffraction technique offers enhanced sensitivity to layer-resolved lateral thermal transport and provides a powerful platform for probing anisotropic materials \cite{kim2021}, phase-change systems \cite{neumann2019}, localized thermal bottlenecks in emerging memory technologies \cite{deshmukh2022}, and more generally heat transport across buried interface in multilayer device architectures.\\

\noindent \textbf{Characterization of epitaxial GaN film}

\begin{figure}
    \centering
    \includegraphics[width=0.85\linewidth]{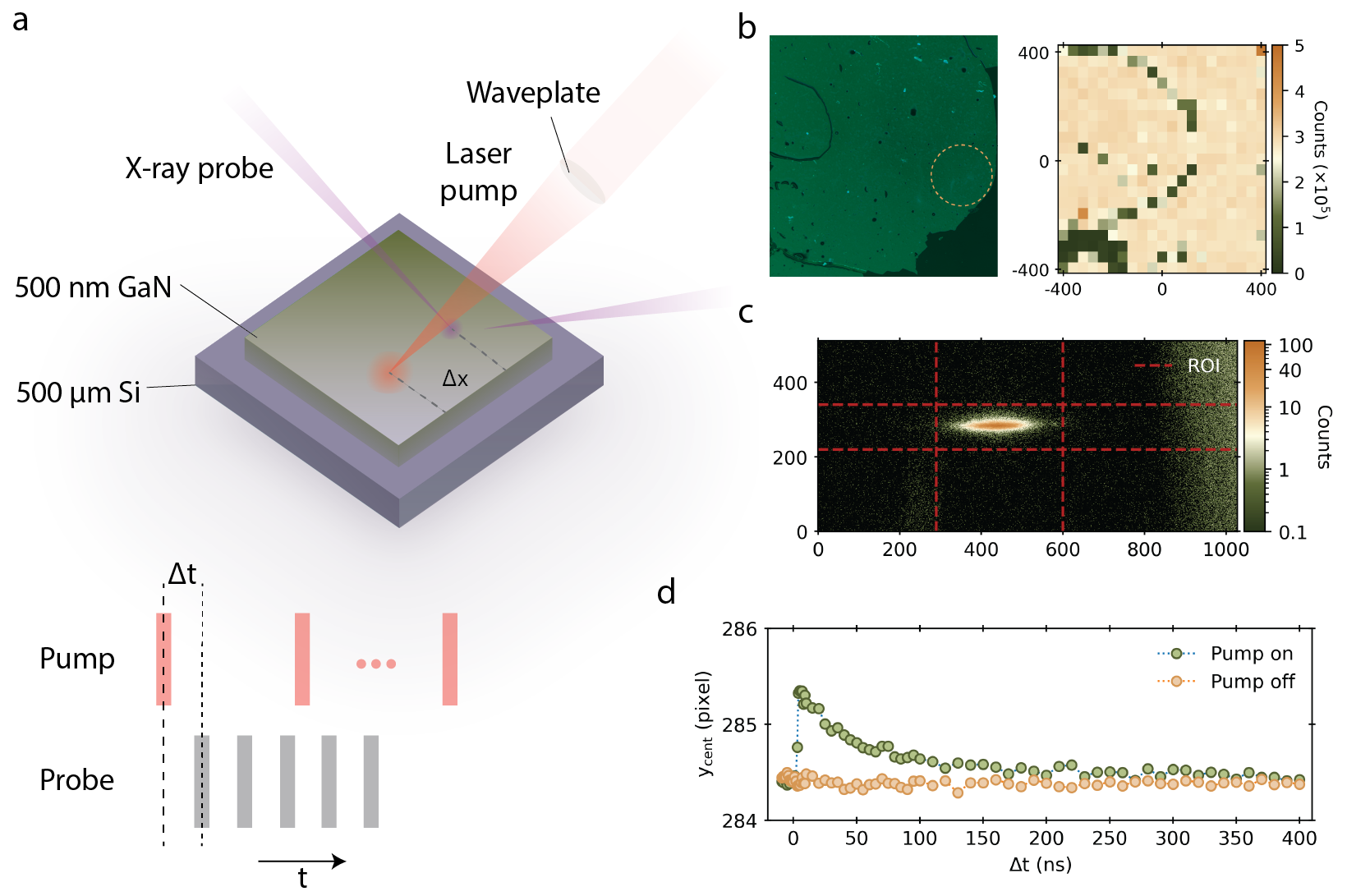}
    \caption{\textbf{GaN film characterization with ultrafast x-ray diffraction.} \textbf{a.} Illustration of the experimental setup. A displaceable 343 nm pump laser at 54 kHz illuminates a location on the epitaxial 500 nm thick GaN thin film on Si substrate. A 11 keV x-ray probe beam (at 6.5 MHz) at a fixed position $\Delta x$ and at a time delay $\Delta t$ relative to the pump laser measures the (002) diffraction peak. \textbf{b.} Optical image of the GaN thin film where the orange dashed-line circle indicates the scan region (diameter of 500 $\mu$m) and x-ray diffraction mapping of the (002) Bragg peak intensity at the wrinkle-feature in the top-left portion of the optical image (length unit in $\mu$m). \textbf{c.} Detector image of the (002) diffraction peak with the length unit in pixels. Red dashed lines demarcate the region of interest to calculate the centroid position of the peak. \textbf{d.} Double-gated measurement of the (002) Bragg peak centroid (y-value) as a function of pump-probe time delay $\Delta t$ at spatial coincidence ($\Delta x = 0$) with the pump on (green) and off (orange).}
    \label{fig:1}
\end{figure}

The sample consists of a 500 nm GaN thin film epitaxially grown on monolayer amorphous boron nitride (aBN) on a GaN substrate under a nitrogen-rich environment using molecular beam epitaxy \cite{kim2023}. The thin film is exfoliated by a thermal release tape and bonded onto a 500 $\mu$m thick silicon substrate by applying heat and pressure to ensure a conformal contact and subsequently annealed for a direct van der Waals bonding (details on this process can be found in Methods). Fig. \ref{fig:1}b presents an optical image of our GaN-on-Si sample, in which wrinkles and residues (black features) are observed as a result of the etching step during the remote epitaxy-based transfer process described in Ref.~\cite{kim2017remote}.
Additional images of the sample can be found in the Supplementary Information S1. Ultrafast x-ray diffraction measurements in a reflection geometry were performed at Beamline 7-ID-C at the Advanced Photon Source at Argonne National Laboratory. Fig. \ref{fig:1}a displays the setup with a 343 nm optical pump laser with pulse duration of 400 fs, spot size of $\sim$13 $\mu$m, and repetition rate of 54 kHz. The photon energy of the optical laser is above the bandgap of GaN. The x-ray probe is focused down to a 1 $\mu$m spot-size and monochromatized to 11 keV with a repetition rate of 6.5 MHz. The spatial separation between the pump and probe beams can be controlled with a resolution of 1 $\mu$m using a piezostage. The fluence of the optical pump laser is changed using an optical waveplate and a polarizer. We nominally set the waveplate angle to 55$^\circ$ such that the fluence level of 3 mJ/cm$^2$ does not lead to sample degradation.

We focus on measuring the strong (002) Bragg peak of GaN ($2\theta = 23.8^\circ$) using a two-dimensional pixel array detector as shown by the peak intensity map in Fig. \ref{fig:1}c. A spatial mapping of the Bragg peak integrated intensity scanned over a relatively smooth area of the film is taken in Fig. \ref{fig:1}b and reveals wrinkle-like defects that match the morphology of the SEM image taken on the left of Fig. \ref{fig:1}b. On the detector image, we delimit a region of interest (shown in red dashed lines) to calculate the centroid of the peak. With the optical pump laser and the x-ray probe positioned at the same location on the film (i.e. the in-plane displacement from the pump to the probe is $\Delta x = 0$), time series measurements were taken as a function of pump-probe time delay ($\Delta t$). Measurements were also taken at negative $\Delta t$ to obtain a baseline for the pre-photoexcitation signal. At a constant fluence, Fig. \ref{fig:1}d shows an increase of the (002) diffraction peak $y$-centroid upon photoexcitation with the optical pump, which corresponds to an increase in the lattice constant of the thin film from dynamic photoinduced strain. 

For consistency throughout the text, we plot the angular shift as a positive value to indicate an increase in the lattice constant. Measurements are performed in a ``double-gate mode'' in which the difference between images taken both before (orange) and after (green) the arrival of the optical pump serves to improve the signal-to-noise and remove electron-refilling-related artifacts (a comparison with data of the single-gated mode is presented in the Supplementary Information S2).

As the pump-probe time delay $\Delta t$ increases, the centroid shift gradually decreases with a global time constant of $\sim$60 ns and eventually reaches its pre-excitation value beyond 400 ns (maximum $\Delta t$ taken during the experiment). The centroid shift drops more rapidly as a function of $\Delta t$ at early timescales ($\Delta t$ < 20 ns) in comparison to the tail of the signal. The integrated intensity of the Bragg peak does not change significantly upon photoexcitation, varying only by a few percent. By contrast, the peak full width at half maximum (FWHM) does show a change after the arrival of the pump pulse, indicating a transient inhomogeneous spatial stress and with a pump-probe time delay dependence that is characterized by a smaller time constant of $\sim$20 ns (details in Supplementary Information S3), consistent with other reports on semiconductor or oxide heterostructures \cite{lee2019,nyby2020,jo2021,lee2023}. The FWHM response accompanying the sharp centroid shift at early timescales supports the idea that two timescales are involved: a fit to the signal with a double-exponential model improves over one with a single-exponential (shown in Supplementary Information S4). The creation of photogenerated charge carriers (electronic strain) or propagation/generation of ballistic phonons may induce an additional strain to the thin film with a faster thermalization ($\Delta t$ < 20 ns) in contrast to the diffusive-like relaxation of the lattice towards equilibrium at later timescales ($\Delta t$ > 60 ns). This early time scale is pronounced in the cross-plane direction where the characteristic thermal redistribution time is estimated to be 8 ns based on the thickness of the film (Supplementary Information S9). By considering the characteristic time and length scales involved in our GaN-on-Si structure, the thermal conductivity of the film can be calculated from a simple model based on the phonon diffusive regime of the lattice relaxation.\\

\noindent \textbf{Extraction of the thermal conductivity with planar sensitivity}

\begin{figure}
    \centering
    \includegraphics[width=0.45\linewidth]{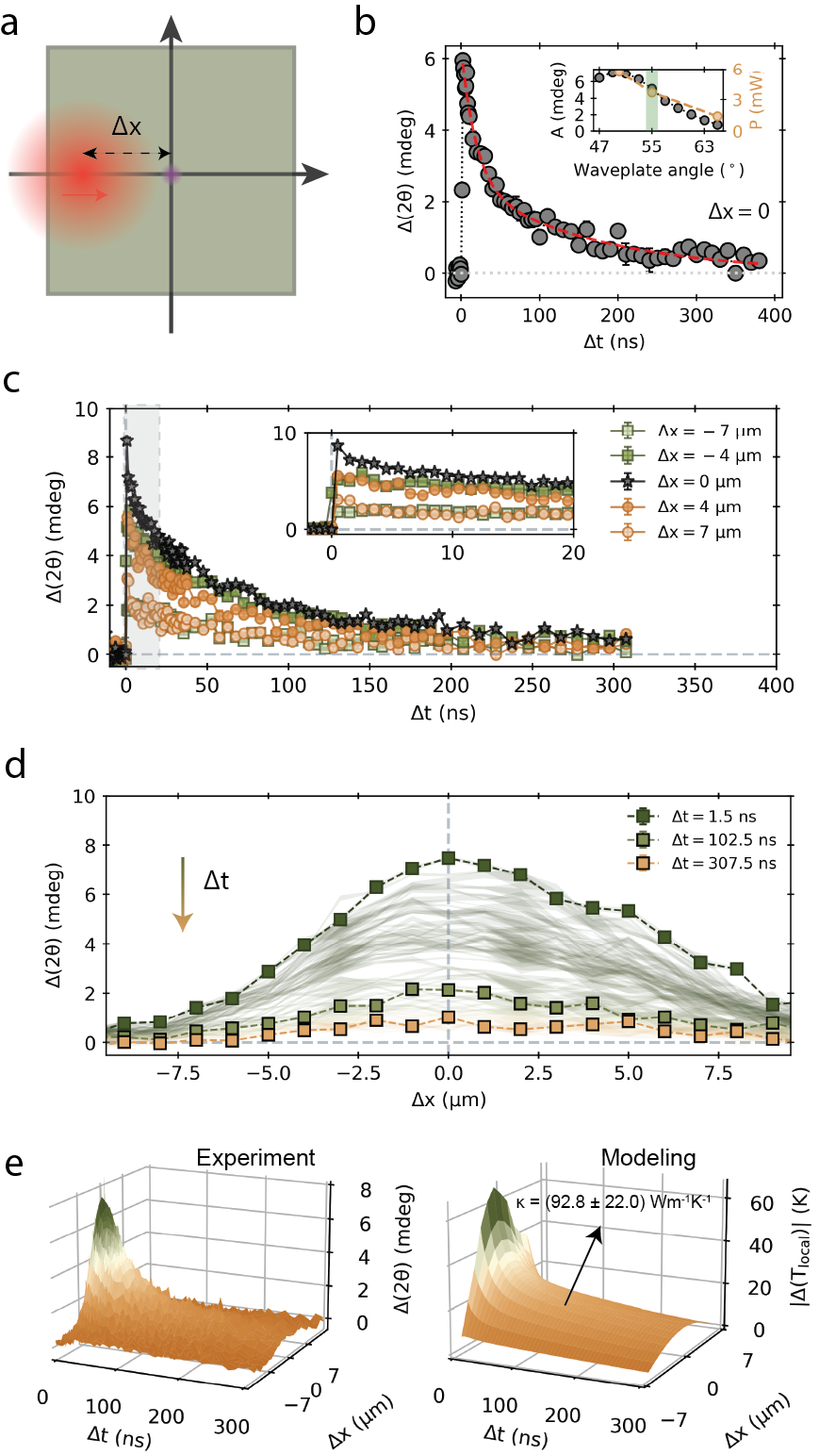}
    \caption{\textbf{Determination of the lateral thermal conductivity.} \textbf{a.} Illustration of the measurement scheme. The x-ray probe is located at the origin. Time series measurements are taken as the pump laser is scanned horizontally along $\Delta y = 0$. \textbf{b.} $2\theta$ angular shift at spatial coincidence ($\Delta x = 0$) as a function of pump-probe time delay $\Delta t$. Inset shows the fitted maximum change in $2\theta$, $A$, immediately after the pump and the measured power as a function of waveplate angle. All subsequent measurements are performed at a waveplate angle of 55$^\circ$. \textbf{c.} Time series measurements of $\Delta(2\theta)$ at different spatial separations between pump and probe. Inset is an enlarged view at small time delays. \textbf{d.} Measurements of $2\theta$ difference at increasing pump-probe time delays (green to orange) versus the relative pump-probe spatial separation. Three time delays are selected as representative scans at early (1.5 ns), intermediate (102.5 ns), and late (307.5 ns) times. \textbf{e}. Volumetric plot of the measured data (left) and heat conduction modeling (right) to extract the local temperature, and by extension, the thermal conductivity through a thermal transport model.}
    \label{fig:2}
\end{figure}

To emphasize the lattice strain, we convert the centroid shift to a $2\theta$ shift of the Bragg peak using the pixel size of the detector array and the detector-sample distance and plot this $2\theta$ difference as a function of $\Delta t$ in Fig. \ref{fig:2}b. At a constant fluence of 3 mJ/cm$^2$ and at spatial coincidence between pump and probe beam spots, the maximum $2\theta$ shift is 6 mdeg, which is equivalent to a $c$-lattice constant change of around 130 fm (or a strain of 2.5$\times 10^{-4}$ with the signal noise levels determined between $10^{-6}$ to $10^{-7}$). The maximum $2\theta$ shift initially increases with the fluence, but plateaus at a certain level due to permanent sample degradation (inset of Fig. \ref{fig:2}b). We choose to perform the rest of the experiment at a fluence level (3 mJ/cm$^2$) below this saturation, indicated by the green region. More details of the fluence dependence are shown in Supplementary Information S5.

To obtain information about the lateral thermal transport of our thin film, we perform measurements where we spatially separate the optical pump laser and the x-ray probe spots on the thin film, with a spatial resolution of less than a micrometer. We choose to measure the diffraction in a spatial region that is relatively uniform and free of defects. As illustrated in the scheme of Fig. \ref{fig:2}a, for convenience, we fix the x-ray probe at the center of our region and gradually scan the optical pump in different horizontal positions from left to right relative to the x-ray probe in steps of 1 $\mu$m, including at spatial coincidence. At each position, we perform several time series measurements from $\Delta t = -10$ ns to $\Delta t = 400$ ns with a higher density of points at early times and sum the counts to improve the signal-to-noise statistics. As shown in Supplementary Information S6, the signal can be resolved up to a distance of $\Delta x = \pm 13\ \mu$m between the pump and probe with decent counting rates. To avoid the confusion, we emphasize that the $\Delta x$ here is only the relative displacement, but not a vector along the x-axis as used in the heat conduction equation modeling. We note that the signal of the spatially resolved angular shift does not match the profile of the laser intensity, and we also take measurements where we scan the positions in the orthogonal direction to verify the isotropy of the laser-induced heat which displays similar results (shown in Supplementary Information S7). 

Fig. \ref{fig:2}c plots the angular shift as a function of pump-probe time delay for a subset of $\Delta x$ values (0, $\pm$4 and $\pm$7 $\mu$m, where negative indicates to the left and positive, to the right). The inset provides an enlarged view at small time delays. The signal between equivalent positions left and right of the probe coincide with each other, reflecting the isotropy of the heat signal relative to the center. Furthermore, as the distance between the pump laser and the x-ray probe increases, the angular shift decreases at a particular $\Delta t$ at all measurable time delays. We present an equivalent view of the data in Fig. \ref{fig:2}d by plotting the angular shift as a function of pump-probe distance at every measured time delay $\Delta t$. The symmetry between the left and right parts of Fig. \ref{fig:2}d is apparent (with the slight asymmetry due to error in the spatial resolution between pump and probe) in addition to a gradual decline of the angular shift signal at all $\Delta x$ with increasing time. As the angular shift decreases with both pump-probe time delay $\Delta t$ and pump-probe distance $\Delta x$, it reflects a heat transport equation that describes the heat dissipation in the thin film. As mentioned previously, the early time scale data ($\Delta t < 20$ ns) contain a sharp decrease of the angular shift followed by a second component near the tail of the measurement ($\Delta t > 60$ ns) that resembles diffusive heat transport. This can be observed in Fig. \ref{fig:2}d as a prominent slender hump at early times on top of an overall parabolic-shape of the diffusive profile. 

To model the experimental data and extract values of the thin film thermal conductivity and the GaN-Si TBC, we build a thermal transport model based on the heat conduction equation as described in detail in Supplementary Information S8 and S9. We convert the angular shift data into a local temperature $\Delta T$ through its relation involving the linear thermal expansion coefficient $\cot(\theta)\Delta\theta = -\alpha\Delta T$. The largest change in $2\theta$ immediately after the optical pump and at spatial coincidence corresponds to a change in local temperature of around 70 K. We employ a three-dimensional generalized heat conduction equation on the local effective temperature $T(\mathbf{r}, t)$ (where $\mathbf{r}$ is the local coordinate) shown in Fig. \ref{fig:2}e as
\begin{equation}
C \rho \frac{\partial T(\mathbf{r}, t)}{\partial t} = \nabla \cdot \left(\overleftrightarrow{\mathbf{k}} \nabla T(\mathbf{r}, t)\right),
\end{equation}
where $\overleftrightarrow{\mathbf{k}}$ is the thermal conductivity tensor of the thin film, assumed to be diagonal, with in-plane components $k_{\parallel} = k_{xx} = k_{yy}$ and a cross-plane component $k_{\perp} = k_{zz}$ (these components may vary spatially), $C$ is the heat capacity, and $\rho$ is the mass density. We perform numerical calculations of the aforementioned model using the finite difference method (details in Supplementary Information S11). To account for the early time electronic strain contribution, we do not take into account the signal measured before 8 ns when setting the initial temperature profile in our procedure such that the thin film is given sufficient time to thermally redistribute. In order to obtain a lateral thermal conductivity value in our three-dimensional model, we fix the cross-plane thermal conductivity to the value obtained in our FDTR experiment (with a value of $k_{\perp}=$ 65 W/m$\cdot$K, shown later on). By using our finite difference method, we perform a one-parameter fit of the comprehensive $\Delta T$ dataset by minimizing the loss function to obtain a lateral thermal conductivity value of GaN (on 500-$\mu$m Si) of $k_{\parallel} =$ (92.8 $\pm$ 22.0) W/m$\cdot$K. This value is much lower than that of bulk GaN of around 230-250 W/m$\cdot$K, likely due to a classical size effect \cite{chen2005} from the film thickness relative to the phonon mean free path distribution and the interfacial effect of GaN-Si resulting in a sizeable TBC. We plot the three-dimensional projection of our numerical calculation of the local temperature mapping as a function of $\Delta x$ and $\Delta t$ in Fig. \ref{fig:2}e which agrees well with the experimental data.

Furthermore, from the ultrafast x-ray diffraction data, we can obtain an estimate for the TBC of the GaN/Si interface as shown in Ref. \cite{nyby2020} from the expression
\begin{equation}
    \tau = \frac{C\rho l}{G}
\end{equation}
where $\tau$ is the relaxation time constant associated with the thermal transport, $C$ is the specific heat capacity of GaN, $\rho$ is the mass density, $l$ is the thickness of the film, and $G$ is the GaN/Si TBC. Unlike the calculation of the thermal conductivity in which we identified two distinct contributions to the angular shift with differing time constants, here we assume that both entities (electronic or other induced strain contribution and diffusive phonon contribution) are involved in the thermal transport through the interface. As such, we use a single-exponential model to fit for an effective $\tau$ based on the angular shift data at $\Delta x = 0$ from which we obtain a value of $\tau = (45 \pm 3)$ ns. By using $C = 35.4$ J/mol$\cdot$K, $\rho = 6.08$ g/cm$^{3}$, and $l = 500$ nm, we obtain a GaN/Si TBC value of $G = (2.8 \pm 0.2)\times10^7$ W/m$^2\cdot$K, consistent with previous literature \cite{sarua2007,cho2014}.\\

\noindent \textbf{Wrinkle-affected local thermal transport}

\begin{figure}
    \centering
    \includegraphics[width=0.95\linewidth]{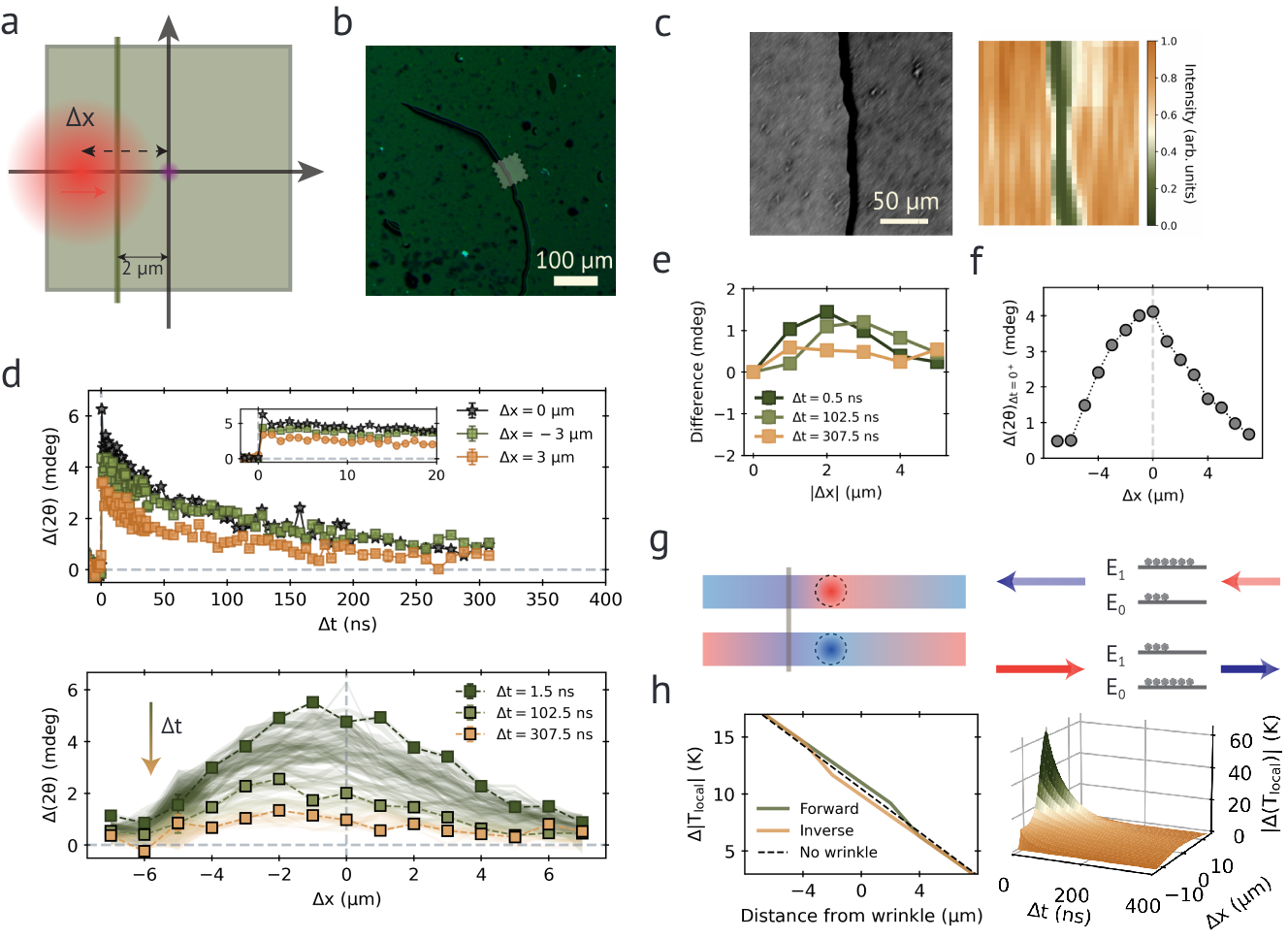}
    \caption{\textbf{Wrinkle-affected asymmetric local thermal transport.} \textbf{a.} Schematic of the measurement scheme. The x-ray probe is located at the origin and the vertical wrinkle is at $\Delta x = -2$ $\mu$m. Time series measurements are taken as the pump laser is scanned horizontally along  $\Delta y = 0$. \textbf{b.} Optical image of the GaN thin film; light-colored area indicates the scan region. Scale bar is 100 $\mu$m. \textbf{c.} Images of the discernible $\sim1\ \mu$m-wide wrinkle with SEM (left) and scanning XRD (right). Scale bars are 50 $\mu$m and 1 $\mu$m/pixel, respectively. \textbf{d.} (top) Time series measurements of $2\theta$ shift versus pump-probe time delay with the pump laser located at spatial coincidence with the probe (black), to the left of the wrinkle (green), and to the right of the wrinkle (orange). Inset is an enlarged version at small time delays. (bottom) Measurements of $2\theta$ shift at increasing pump-probe time delays (green to orange) versus the relative pump-probe position. \textbf{e.} Left-right asymmetry of the $2\theta$ difference plotted versus the relative distance between the pump and probe lasers. \textbf{f.} Fitted value for the maximum $2\theta$ shift immediately after the pump pulse as a function of relative pump-probe position. \textbf{g.} Illustration of the wrinkle-affected thermal transport in the thin film and the wrinkle-induced potential as a two-level model. \textbf{h.} Plot of the local temperature change in the forward and reverse directions relative to the wrinkle and local temperature profile extracted from the wrinkle-affected measurement series.}
    \label{fig:3}
\end{figure}

As thin film transfer processes onto substrates typically involve chemical etchants and stressor materials, it is common to have residual defects on the film with local imperfections such as cracks and wrinkles. The GaN-on-Si sample is no exception as revealed by the presence of a discernible $\sim1\ \mu$m-wide wrinkle revealed in SEM and x-ray diffraction mapping images (Figs. \ref{fig:3}b and \ref{fig:3}c). We investigate how heat induced by the optical pump laser travels through the wrinkle enabled by our ultrafast x-ray diffraction technique. Furthermore, this provides a microscopic picture of the heat transport near this localized defect through our mapping of the local temperature $T$. We proceed by performing a measurement scheme similar to the one used to extract the thermal conductivity, but here we fix the position of the x-ray probe spot to 2 $\mu$m right to a locally straight line wrinkle (Fig. \ref{fig:3}a). Afterwards, as done previously, we perform time series measurements when scanning the pump laser at different 1 $\mu$m step positions along a line perpendicular to the $\sim1\ \mu$m-wide wrinkle, while passing over the wrinkle ($\Delta x = -2\ \mu$m) and the x-ray probe ($\Delta x = 0 \mu$m).

The upper plot of Fig. \ref{fig:3}d displays measurements of the angular shift as a function of pump-probe time delay when the pump is at a spatial coincidence with the probe (black), 3 $\mu$m (green), and -3 $\mu$m (orange). There is a stark asymmetry in the signal between equidistant pump distances due to the presence of the wrinkle. As the wrinkle defect likely decreases the local thermal conductivity of the thin film, as shown previously in diamond \cite{sood2018} and InN \cite{sun2019} films, the $2\theta$ angular shift is more pronounced on the negative direction (where the wrinkle is) compared to the positive direction as the laser-induced heat travels less effectively through the wrinkle and the lattice strain remains at a higher level immediately after the pulse when $\Delta t$ is a few nanoseconds. The bottom plot of Fig. \ref{fig:3}d demonstrates this observation more explicitly with a clear asymmetry between the negative and positive sides of the plot, in contrast to the situation of the pristine region of the film in Fig. \ref{fig:2}d.

We accentuate this asymmetry by plotting the difference between the angular shifts at equidistant between left and right (Fig. \ref{fig:3}e) and the maximum angular shift immediately after the optical pump as a function of pump-probe distance $\Delta x$ (Fig. \ref{fig:3}f). Immediately after the pulse, the difference is largest at the position of the $\sim1\ \mu$m-wide wrinkle ($|\Delta x| = 2\ \mu$m) and gradually decreases with $\Delta t$ while shifting towards larger $|\Delta x|$, indicating that the asymmetry may be tied to the lateral diffusion of the heat pulse in the thin film. The maximum angular shift shows a strong asymmetry immediately after the pulse indicating that the initial heat pulse after a few nanoseconds is tied to the presence of the wrinkle. We do not probe for timescales less than a nanosecond, therefore information about wrinkle-induced coherent phonon propagation or attenuation effects due to thermal motion are beyond this current study. Our study reveals how this wrinkle-affected thermal transport is asymmetric in nature as heat is able to flow efficiently in one direction compared to the other due to a sharp drop in thermal conductivity at the wrinkle. As shown in Supplementary Information S10, we can model this situation by a two-level system to describe the thermal excitation of the wrinkle (Fig. \ref{fig:3}g) \cite{anderson1972anomalous}. Calculations of the local temperature as a function of distance to the wrinkle in both forward and reverse directions under steady-state conditions reveal this asymmetric transport (Fig. \ref{fig:3}h, left). To extract the local thermal conductivity of the wrinkle, we use the same thermal model as the pristine case, but allow for the thermal conductivity to be locally perturbed as $k_w = k_{\parallel, \text{hom}} + \Delta k_w$, where $k_{\parallel, \text{hom}}$ is the assumed homogeneous conductivity along the in-plane direction. Using a similar numerical calculation (Fig. \ref{fig:3}h, right), we obtain $k_w = (21.2 \pm 1.2)$ W/m$\cdot$K, which represents a 4 to 5 fold decrease of the thermal conductivity. Furthermore, we obtain a GaN-Si TBC value near the wrinkle defect of $G_w = (2.12 \pm 0.05)\times 10^7$ W/m$^2\cdot$K, which corresponds to a 25\% decrease compared to the value calculated in the pristine case and from literature \cite{sarua2007,cho2014}. Similar to diamond (2.5-fold decrease) \cite{sood2018} and InN (10-fold decrease) \cite{sun2019}, our direct mapping visualization of the thermal conductivity suppression localized at the wrinkle defect may have far-reaching implications for models of heat dissipation in nanoscale devices, which can be established based on results for high-quality pristine thin films. This decrease of the thermal conductivity and of the TBC near the wrinkle may lead to local hotspots, which would decrease the lifetime and reliability of devices.\\

\noindent \textbf{Comparison with FDTR and literature benchmarking}

\begin{figure}
    \centering
    \includegraphics[width=0.8\linewidth]{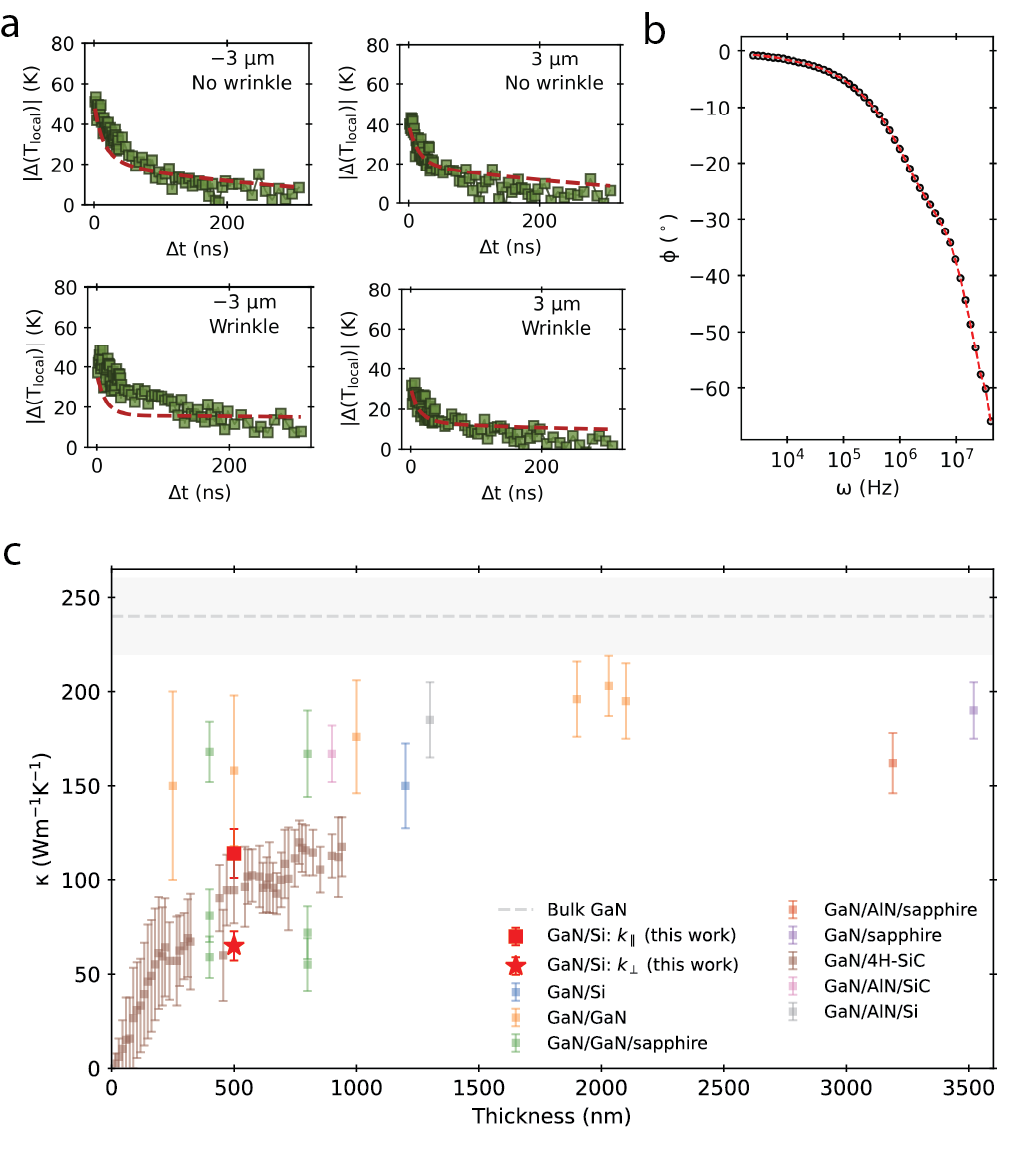}
    \caption{\textbf{Diffraction-based thermal metrology for thermal conductivity and literature comparison.} \textbf{a.} Computational modeling of the thermal model (red dashed lines) plotted on the experimental x-ray data (green squares) in the absence (top) and presence (bottom) of the wrinkle to extract the lateral thermal conductivity at a specific pump-probe beam distance. \textbf{b.} Frequency-domain thermoreflectance (FDTR) measurements on the GaN thin film with an Au transducer layer. The red dashed line corresponds to a fit of the data to extract the cross-plane thermal conductivity of GaN and the thermal boundary conductances associated with the Au/GaN and GaN/Si interfaces. \textbf{c.} Comparison among reported values of the room temperature thermal conductivity against thickness of GaN thin film and bulk values in literature, including those obtained in this work.}
    \label{fig:4}
\end{figure}

To extract quantitative values of thermal conductivity and TBC from the spatially resolved ultrafast x-ray diffraction measurements, we perform a fit of the entire time series dataset at each pump-probe positional configuration shown in Fig. \ref{fig:4}a, with or without the presence of defects, using a simple heat conduction model. During the modeling procedure, we subtract off the initial strain profile with a small time constant and use the predominantly diffusive profile to obtain the value of thermal conductivity for GaN. The size of error of the fitted thermal conductivity is comparable to other experimental techniques. Due to the smaller value of TBC between GaN and Si (an order of magnitude less than typical interfaces) and the short light penetration depth of GaN, most of the heat lingers in the thin film such that the one-dimensional heat conduction modeling in the in-plane direction would adequately describe the experiment. Nevertheless, we expect that additional sensitivity in the planar direction is achieved from the beam offset such that the signal picks up both in-plane and cross-plane contributions, which are comprehensively captured in our three-dimensional model. The sensitivity to both directional components can account for the discrepancies in thermal conductivity values obtained between our x-ray experiment and those from an FDTR measurement presented below. Due to its capability of extracting lateral thermal conductivity values, our methodology based on beam offset ultrafast diffraction may serve as a complementary non-contact, non-destructive measurement to traditional techniques, such as TDTR, which are mainly sensitive to the cross-plane thermal conductivity.

To corroborate our values of GaN thin film thermal conductivity and provide a comparative baseline, we perform FDTR measurements on a portion of the same sample on top of which a 100 nm layer of gold serves as a transducer layer. We perform the measurement at different positions on the sample, which provides values of probe phase delay versus frequency of the pump laser shown in Fig. \ref{fig:4}b. We fit the FDTR phase data using a three-layer isotropic heat conduction model (details in Methods and Supplementary Information S12) to obtain a (cross-plane) thermal conductivity of GaN of (65 $\pm$ 8) W/m$\cdot$K and a TBC between the GaN and Si of (2.82 $\pm$ 0.45)$\times 10^7$ W/m$^2\cdot$K. The GaN-Si TBC value matches the value obtained from our ultrafast x-ray diffraction measurement, and both values are consistent with previous reports of GaN thin films on Si of $\sim$3$\times10^7$ W/m$^2\cdot$K \cite{sarua2007,cho2014}. The lateral thermal conductivity value obtained in the x-ray experiment is approximately 1.5 times larger than that measured from the FDTR experiment, which is assumed to be purely cross-plane. This may be expected due to the wurzite crystal structure of GaN \cite{wu2016,tran2023} and the substrate support: the anisotropy factor is $\eta = 2(k_{\parallel}-k_{\perp})/(k_{\parallel}+k_{\perp}) = 35\%$. In Fig. \ref{fig:4}c, we compare these values with literature of GaN thin films on other substrates and of various thicknesses with the upper limit from bulk crystals (more details and references in Supplementary Information S13). The values from the x-ray measurement follow the trend of increasing thermal conductivity with increasing thickness as shown by previous studies of the size effect \cite{beechem2016,ziade2017}. The value of $k_{\perp}$ from FDTR is slightly lower than that of GaN/SiC due to the smaller lattice mismatch and that of GaN/diamond due to the higher conductivity of the substrate. Moreover, our value of lateral $k_{\parallel}$ lies on the trend curve of GaN/SiC and may appear to be on par with the trend from the previous GaN/Si measurement of Ref. \cite{sarua2007} taken in a much thicker film (blue square in Fig. \ref{fig:4}c). In contrast to SiC, diamond or GaN substrates, the thermal conductivity of GaN thin films on Si remains competitive for future power electronics while being a cost-effective option. \\

\noindent \textbf{Discussion}

In this work, we demonstrate that spatially resolved ultrafast x-ray diffraction enables direct, non-contact access to in-plane thermal transport and interfacial boundary conductance in thin film systems. The combination of beam offset geometry and deep x-ray penetration provides unique sensitivity to lateral heat conduction and localized thermal variations, including the influence of defect features that are difficult to access with conventional optical techniques.

Beyond quantitative extraction of thermal properties, this approach offers practical advantages: it requires no transducer layer, preserves the sample integrity, and is compatible with buried or encapsulated structures. These attributes make it well-suited for probing anisotropic transport, evaluating local disruptions in thermal flow, and studying device-relevant materials where thermal gradients are highly directional.

As a result, this work demonstrates that ultrafast x-ray diffraction establishes itself as a powerful and broadly applicable platform for characterizing heat transport in emerging materials systems, including van der Waals heterostructures, phase-change media, and neuromorphic device architectures.

\bibliographystyle{apsrev4-1.bst}
\bibliography{references.bib}

\vspace{1cm}

\textbf{Author contributions}: TN and ML conceived the experiments and ML supervised the project. TN, CF, MC, TEE, ZC, AC, DCC, ER, and ML performed the effective modeling and analysis of the x-ray diffraction data. KQ and JK grew and transferred the epitaxial GaN sample on SiO$_2$ substrate. TN, KN, ZC, TL, DAW, and HW performed the pump-probe time-resolved x-ray diffraction experiment at Beamline 7-ID-C at Argonne National Laboratory. TN, TEE, SNK, and DS performed the sample characterization by microscopy. BL and TN under the guidance of SVB and GC performed the FDTR measurements. TN, CF and ML wrote the manuscript with input from all other authors.\\

\textbf{Acknowledgements}: We gratefully thank Ching-Tzu Chen, Keith Nelson, Nguyen Tuan Hung, Leonid Zhigilei, and Jing Kong for extensive discussions on thermal transport in thin film materials. TN, CF, AC, and ML acknowledge the support from U.S. Department of Energy, Office of Science, Basic Energy Sciences Award No. DE-SC0020148. TN is supported by the National Science Foundation Designing Materials to Revolutionize and Engineer our Future Program (NSF DMREF) under award No. DMR-2118448 and by the 2023-2024 MIT School of Engineering Distinguished Energy Efficiency Fellowship. BL and SVB are supported by U.S. Department of Energy, Office of Science, Basic Energy Sciences Award DE-FG02-02ER45977. ML thanks the support from Class of 1947 Career Development Professor Chair and support from Dr. R. Wachnik. HW acknowledges the support for instrumentation development by U.S. Department of Energy (DOE), Office of Science (SC), Basic Energy Sciences (BES), Materials Sciences and Engineering Division. KN and HW acknowledge the support for assisting the data collection by U.S. Department of Energy, Office of Science, National Quantum Information Science Research Centers. The x-ray scattering experiments used resources of the Advanced Photon Source, a U.S. Department of Energy, Office of Science, User Facility operated by Argonne National Laboratory under Contract No. DE-AC02-06CH11357. This work was performed in part in the MIT Materials Research Laboratory (MRL) Shared Experimental Facilities supported by Division of Materials Research (DMR) Award No. 1419807.\\

\textbf{Competing financial interests}: The authors declare no competing financial interests.

\clearpage

\noindent \textbf{Materials and Methods}\\
\noindent \textbf{Synthesis of gallium nitride thin film}

A layer of 5 $\mu$m GaN is grown by metal-organic chemical vapour deposition (MOCVD) on sapphire to be used as the growth substrate which is cleaned with acetone, isopropyl alcohol, and dilute hydrochloric acid before being loaded in a Veeco Gen200 plasma-assisted molecular beam epitaxy (MBE) system. The sample is outgassed at 700$^\circ$C for 5 minutes to remove the oxide layer, then decreased to 680$^\circ$C. A high temperature effusion boron cell and plasma-assisted nitrogen are used as the sources of boron and nitrogen, respectively, for the growth on aBN, which is performed under a nitrogen-rich environment for 20 minutes at a radiofrequency plasma power of 250 W. The GaN epitaxial layer is grown on the aBN in the same chamber using a gallium effusion cell and plasma-assisted nitrogen at a temperature of 700$^\circ$C and plasma power of 500 W. \\

\noindent \textbf{Transfer and bonding of epitaxial gallium nitride onto silicon wafer}

For the exfoliation of the GaN epitaxial layer, a titanium adhesion layer is deposited by electron-beam deposition, followed by a 1 $\mu$m Ni stressor layer by DC sputtering. Thermal release tape is used to mechanically exfoliate the GaN thin film off the aBN layer. The thermal release tape with the GaN layer is placed on the Si wafer and heated on a hotplate at 120$^\circ$C and pressure is applied on the stack. An annealing step takes place at high temperature under an inert gas environment such that GaN is bonded to the Si wafer via a van der Waals bonding. The GaN film has a homogeneous contact with the Si wafer.\\

\noindent \textbf{Characterization of the gallium nitride thin film}

Optical images were taken using a Zeiss Axioplan 2 optical microscope. Surface morphologies of the GaN film were analyzed using a Zeiss Merlin high-resolution SEM system.\\

\noindent \textbf{Ultrafast x-ray diffraction measurements}

Ultrafast x-ray diffraction experiments were performed at the 7-ID-C beamline of the Advanced Photon Source at Argonne National Laboratory. A 400 fs, 343 nm (3.625 eV) optical pump pulse was created from the frequency-doubling of an amplified Pharos laser system at a repetition rate of 54 kHz. The photon energy of the pump pulse is above the band gap of 3.4 eV. The x-ray probe was monochromatized to an energy of 11 keV with a pulse width of 100 ps at a repetition rate of 6.5 MHz. The GaN-on-Si sample was mounted on a six-circle Huber diffractometer and experiments were performed at room temperature. The optical pump spot size was around 13.3 $\mu$m and the x-ray probe spot size was around 1 $\mu$m at full-width-half-maximum. The measurements were performed at the (002) Bragg peak located at $2\theta = 23.8^\circ$ and diffracted x-ray photons were collected by a two-dimensional pixel array Eiger2 (Dectris) detector with a pixel size of 75 $\mu$m at a distance of 700 mm from the sample. The fluence of the optical pump laser was tunable by rotation of the waveplate which we set to 55$^\circ$ for the rest of the experiment such that the GaN film did not get destroyed during the measurement due to high photon intensity.\\

\noindent \textbf{Frequency-domain thermoreflectance measurements}

We characterize the thermal conductivity of the sample using a frequency domain thermoreflectance (FDTR) technique developed in Refs. \cite{schmidt2009,schmidt2010,yang2013,yang2016}. The samples were coated with a 100 nm layer of Au using a DC-magnetron sputter system (Leica EM ACE600). The Au film serves as an optical transducer for the FDTR measurements.

Our FDTR platform employed a continuous wave (CW) pump laser with a wavelength of 488 nm and another CW probe laser with a wavelength of 532 nm. The pump laser was sinusoidally modulated from around 5 kHz to 40 MHz. The 1/e$^2$ diameter of the laser spot on the sample was about 5.0 $\mu$m for the pump laser and 5.4 $\mu$m for the probe laser. The pump was used to heat up the sample sinusoidally while the probe senses the modulated sample surface reflectance, and thereby its temperature, via the thermoreflectance effect of the coated transducer layer. The power of the reflected probe beam was detected by a balanced detector, using a reference beam split from the laser source to minimize the noise due to the laser power fluctuation. The output of the balanced detector was given to a radio-frequency lock-in amplifier. Before the sample measurement, the phase of the pump beam at each modulation frequency was predetermined. The phase lag between the modulated surface temperature and the sinusoidal heating was then recorded as the FDTR signal. A benchmark calibration is performed with a silicon wafer.

Since our GaN sample is prepared on silicon substrate with a thickness of 500 nm, this thickness is large enough compared to the penetration depth at our pumped frequencies yet adequately thin to be treated as semi-infinite. As such, we fit the FDTR phase data using a three-layer (Au-GaN-Si) isotropic heat conduction model with the thermal conductivity of GaN and two thermal interfacial resistances (Au-GaN and GaN-Si) as fitting parameters. While anisotropic thermal conductivity is expected for the GaN sample, the pump and probe laser spots are well aligned without beam offset in FDTR. A one-dimensional transient heat conduction along the out-of-plane direction is assumed in the model, leaving negligible sensitivity on the in-plane thermal conductivity. Therefore, an isotropic thermal conduction model is used.

\noindent \textbf{Heat conduction modeling of thermal transport}

To model the spatiotemporal temperature dynamics in the GaN/Si system, we numerically solve the three-dimensional heat conduction equation under the assumption of diffusive phonon transport. Given the thin film thickness of $500\,\mathrm{nm}$ and phonon mean free path $\sim 100\,\mathrm{nm}$, thermal transport is well described by the diffusive regime. The governing thermal conduction equation is:

\begin{equation}
C \rho \frac{\partial T}{\partial t} = \nabla \cdot (\overleftrightarrow{\mathbf{k}} \nabla T),
\end{equation}

where $T(\mathbf{r}, t)$ is the effective local temperature at position $\mathbf{r} = (x, y, z)$, $\rho$ is the mass density, $C$ is the specific heat capacity, and $\overleftrightarrow{\mathbf{k}}$ is the thermal conductivity tensor. Anisotropy is accounted for via diagonal elements $k_{xx}, k_{yy}, k_{zz}$, assumed spatially dependent, while off-diagonal terms are neglected.

To capture interfacial heat resistance, the TBC between GaN and Si is modeled through a Neumann-type boundary condition:

\begin{equation}
\begin{split}
        -k_{zz,\mathrm{GaN}}\frac{\partial T_{\rm GaN,b}}{\partial {z^{+}} } = G_{\rm{GaN,Si}} \left(T_{\rm{Si,b}}-T_{\rm{GaN,b}}\right),\\-k_{zz,\mathrm{Si}}\frac{\partial T_{\rm Si,b}}{\partial {z^{-}} } = G_{\rm{GaN,Si}} \left(T_{\rm{GaN,b}}-T_{\rm{Si,b}}\right).
\end{split}
\end{equation}

$G_{\rm{GaN,Si}}$ is the TBC of the interface between Si and GaN. A subscript $b$ represents the spatial location on the contacting boundary layer between the GaN thin film and the Si substrate. It indicates the position at the bottom of GaN and the top of the Si substrate Si. $T_{\rm Si,b}$ and $T_{\rm{GaN,b}}$ are the temperature profiles at the contacting boundary layer for Si and GaN, respectively.

While the experiments only run along the x-axis, the problem reduces to a one-dimensional problem. To extract the in-plane thermal conductivity and TBC, we define a loss function comparing simulated ($T_{\mathrm{s}}$) and experimental ($T_{\mathrm{m}}$) temperatures over all positions along the x-axis $x_i$ and times $t_j$:

\begin{align}
\mathcal{L} &= \sum_{x_i, t_j} \left| T_{\mathrm{m}}(x_i, t_j) - T_{\mathrm{s}}(x_i, t_j) \right| \nonumber \\
&\quad + \sum_{x_i, t_j} \left| \frac{\partial T_{\mathrm{m}}}{\partial x}(x_i, t_j) - \frac{\partial T_{\mathrm{s}}}{\partial x}(x_i, t_j) \right| \nonumber \\
&\quad + \sum_{x_i, t_j} \left| \frac{\partial T_{\mathrm{m}}}{\partial t}(x_i, t_j) - \frac{\partial T_{\mathrm{s}}}{\partial t}(x_i, t_j) \right| \nonumber \\
&\quad + \sum_{t_j} \left| \left( \max_{x_i} T_{\mathrm{m}}(x_i, t_j) - \min_{x_i} T_{\mathrm{m}}(x_i, t_j) \right) - \left( \max_{x_i} T_{\mathrm{s}}(x_i, t_j) - \min_{x_i} T_{\mathrm{s}}(x_i, t_j) \right) \right|.
\end{align}

To handle wrinkle-affected regions, an additional curvature regularization term is introduced:

\begin{equation}
\mathcal{L}_c = \left| \frac{\partial^2 T_{\mathrm{m}}}{\partial x^2} - \frac{\partial^2 T_{\mathrm{s}}}{\partial x^2} \right|,
\end{equation}

and the total loss function becomes:

\begin{equation}
\mathcal{L}_{\mathrm{tot}}(G_{\mathrm{GaN,Si}}, k_w) = \mathcal{L}_{\mathrm{non\text{-}wrinkle}} + \mathcal{L}_{\mathrm{wrinkle}} + \mathcal{L}_c.
\end{equation}

Optimization is performed by scanning through the parameter space of $G_{\mathrm{GaN,Si}}$ and $k_w$ using gradient-based methods (e.g., L-BFGS) or differential evolution to avoid local minima. More details on relevant modeling and fitting on experimental data are shown in Supplementary Information S9-S11.

\clearpage

\clearpage

\clearpage

\clearpage

\end{document}

%\title{Planar thermal transport mapping of an epitaxial gallium nitride film}
\title{Spatiotemporal Mapping of Anisotropic Thermal Transport in GaN Thin Films via Ultrafast X-ray Diffraction: Supplementary Information}

\affiliation{Department of Nuclear Science and Engineering, Massachusetts Institute of Technology, Cambridge, MA 02139, USA}
\affiliation{Center for Computational Science and Engineering, Massachusetts Institute of Technology, Cambridge, MA 02139, USA}
\affiliation{Department of Materials Science and Engineering, Massachusetts Institute of Technology, Cambridge, MA 02139, USA}
\affiliation{Department of Mechanical Engineering, Massachusetts Institute of Technology, Cambridge, MA 02139, USA}
\affiliation{Department of Physics, Massachusetts Institute of Technology, Cambridge, MA 02139, USA}
\affiliation{Department of Electrical Engineering and Computer Science, Massachusetts Institute of Technology, Cambridge, MA 02139, USA}
\affiliation{Linac Coherent Light Source, SLAC National Accelerator Laboratory, Menlo Park, CA 94025, USA}
\affiliation{Stanford Institute for Materials and Energy Sciences, Stanford University, Stanford, CA 94025, USA}
\affiliation{Materials Science Division, Argonne National Laboratory, Lemont, IL 60439, USA}
\affiliation{MIT Media Lab, Massachusetts Institute of Technology, Cambridge, MA 02139, USA}
\affiliation{Advanced Photon Source, Argonne National Laboratory, Lemont, IL 60439, USA}
\affiliation{These authors contributed equally to this work.}

\author{Thanh Nguyen}
\thanks{Corresponding authors. \href{mailto:ngutt@mit.edu}{ngutt@mit.edu}, \href{mailto:jeehwan@mit.edu}{jeehwan@mit.edu}, \href{mailto:mingda@mit.edu}{mingda@mit.edu}}
\affiliation{Department of Nuclear Science and Engineering, Massachusetts Institute of Technology, Cambridge, MA 02139, USA}
\affiliation{These authors contributed equally to this work.}
\author{Chuliang Fu}
\affiliation{Department of Nuclear Science and Engineering, Massachusetts Institute of Technology, Cambridge, MA 02139, USA}
\affiliation{These authors contributed equally to this work.}
\author{Mouyang Cheng}
\affiliation{Center for Computational Science and Engineering, Massachusetts Institute of Technology, Cambridge, MA 02139, USA}
\affiliation{Department of Materials Science and Engineering, Massachusetts Institute of Technology, Cambridge, MA 02139, USA}
\author{Buxuan Li}
\affiliation{Department of Mechanical Engineering, Massachusetts Institute of Technology, Cambridge, MA 02139, USA}
\author{Tyra E. Espedal}
\affiliation{Department of Physics, Massachusetts Institute of Technology, Cambridge, MA 02139, USA}
\affiliation{Department of Electrical Engineering and Computer Science, Massachusetts Institute of Technology, Cambridge, MA 02139, USA}
\author{Zhantao Chen}
\affiliation{Linac Coherent Light Source, SLAC National Accelerator Laboratory, Menlo Park, CA 94025, USA}
\affiliation{Stanford Institute for Materials and Energy Sciences, Stanford University, Stanford, CA 94025, USA}
\author{Kuan Qiao}
\affiliation{Department of Mechanical Engineering, Massachusetts Institute of Technology, Cambridge, MA 02139, USA}
\author{Kumar Neeraj}
\affiliation{Materials Science Division, Argonne National Laboratory, Lemont, IL 60439, USA}
\author{Abhijatmedhi Chotrattanapituk}
\affiliation{Department of Electrical Engineering and Computer Science, Massachusetts Institute of Technology, Cambridge, MA 02139, USA}
\author{Denisse Cordova Carrizales}
\affiliation{Department of Nuclear Science and Engineering, Massachusetts Institute of Technology, Cambridge, MA 02139, USA}
\author{Eunbi Rha}
\affiliation{Department of Nuclear Science and Engineering, Massachusetts Institute of Technology, Cambridge, MA 02139, USA}
\author{Tongtong Liu}
\affiliation{Department of Physics, Massachusetts Institute of Technology, Cambridge, MA 02139, USA}
\author{Shivam N. Kajale}
\affiliation{MIT Media Lab, Massachusetts Institute of Technology, Cambridge, MA 02139, USA}
\author{Deblina Sarkar}
\affiliation{MIT Media Lab, Massachusetts Institute of Technology, Cambridge, MA 02139, USA}
\author{Donald A. Walko}
\affiliation{Advanced Photon Source, Argonne National Laboratory, Lemont, IL 60439, USA}
\author{Haidan Wen}
\affiliation{Materials Science Division, Argonne National Laboratory, Lemont, IL 60439, USA}
\affiliation{Advanced Photon Source, Argonne National Laboratory, Lemont, IL 60439, USA}
\author{Svetlana V. Boriskina}
\affiliation{Department of Mechanical Engineering, Massachusetts Institute of Technology, Cambridge, MA 02139, USA}
\author{Gang Chen}
\affiliation{Department of Mechanical Engineering, Massachusetts Institute of Technology, Cambridge, MA 02139, USA}
\author{Jeehwan Kim}
\thanks{Corresponding authors. \href{mailto:ngutt@mit.edu}{ngutt@mit.edu}, \href{mailto:jeehwan@mit.edu}{jeehwan@mit.edu}, \href{mailto:mingda@mit.edu}{mingda@mit.edu}}
\affiliation{Department of Mechanical Engineering, Massachusetts Institute of Technology, Cambridge, MA 02139, USA}
\author{Mingda Li}
\thanks{Corresponding authors. \href{mailto:ngutt@mit.edu}{ngutt@mit.edu}, \href{mailto:jeehwan@mit.edu}{jeehwan@mit.edu}, \href{mailto:mingda@mit.edu}{mingda@mit.edu}}
\affiliation{Department of Nuclear Science and Engineering, Massachusetts Institute of Technology, Cambridge, MA 02139, USA}

\date{\today}

\maketitle

\onecolumngrid

\clearpage

\normalsize
\raggedright
This supplementary file contains the following elements:\\
\textbf{Supplementary Text with Sections S1 to S13}\\
\textbf{Figs. S1 to S11}\\
\textbf{Table S1}\\
\tableofcontents

\makeatletter
\renewcommand \thesection{S\@arabic\c@section}
\renewcommand\thetable{S\@arabic\c@table}
\renewcommand \thefigure{S\@arabic\c@figure}
\renewcommand \theequation{S\@arabic\c@equation}
\setcounter{figure}{0}
\makeatother

\clearpage

\justifying

\section{Additional images of the GaN thin film on silicon}
\begin{figure}[h!]
    \centering
    \includegraphics[width=0.45\linewidth]{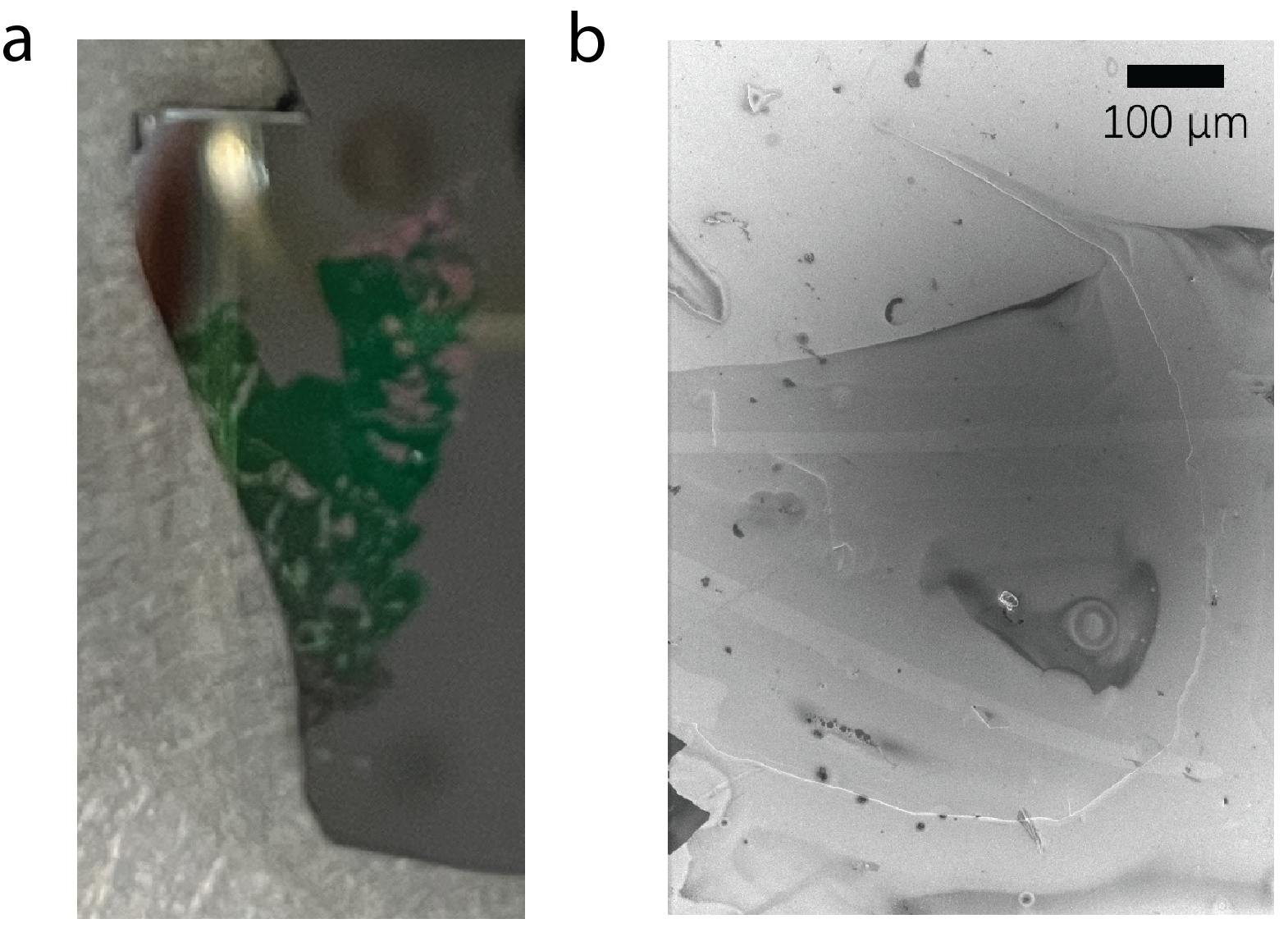}
    \caption{\textbf{a.} Camera image of the GaN thin film (in green) after being transferred onto the piece of silicon substrate using heat and pressure. \textbf{b.} Scanning electron microscope image of the GaN thin film visualizing wrinkles and other local imperfections. The scale bar indicates the length of 100 $\mu$m.}
    \label{fig:S1}
\end{figure}

%\section{Sign of angular shift from centroid shift of Bragg peak}

%\begin{figure}[h!]
%    \centering
    %\includegraphics[width=0.45\linewidth]{FigureS2.png}
%    \caption{Centroid $y$-position of the Bragg peak on the two-dimensional pixel detector in pixels versus $\theta$ in degrees. An increase in the $y$-centroid position corresponds to a decrease in $\theta$, i.e. thermal expansion of the GaN thin film.}
 %   \label{fig:S2}
%\end{figure}

%To verify the sign of the angular shift from the measured centroid shift of the Bragg peak on the two-dimensional pixel detector, we performed a theta scan to obtain a set of images. Upon fitting the Bragg peak position with a two-dimensional Gaussian or using the center-of-mass, the $y$-centroid position shows a decreasing relation with $\theta$ which indicates that an increase in $y$-centroid position (as measured upon photoexcitation) results in decreasing $\theta$ as shown in Fig. \ref{fig:S2}. This corresponds to an increase in lattice constant and thereby, a situation of thermal expansion which is expected due to the positive linear expansion coefficient of GaN.

\section{Single-gated versus double-gated mode acquisition}

\begin{figure}[h!]
    \centering
    \includegraphics[width=0.45\linewidth]{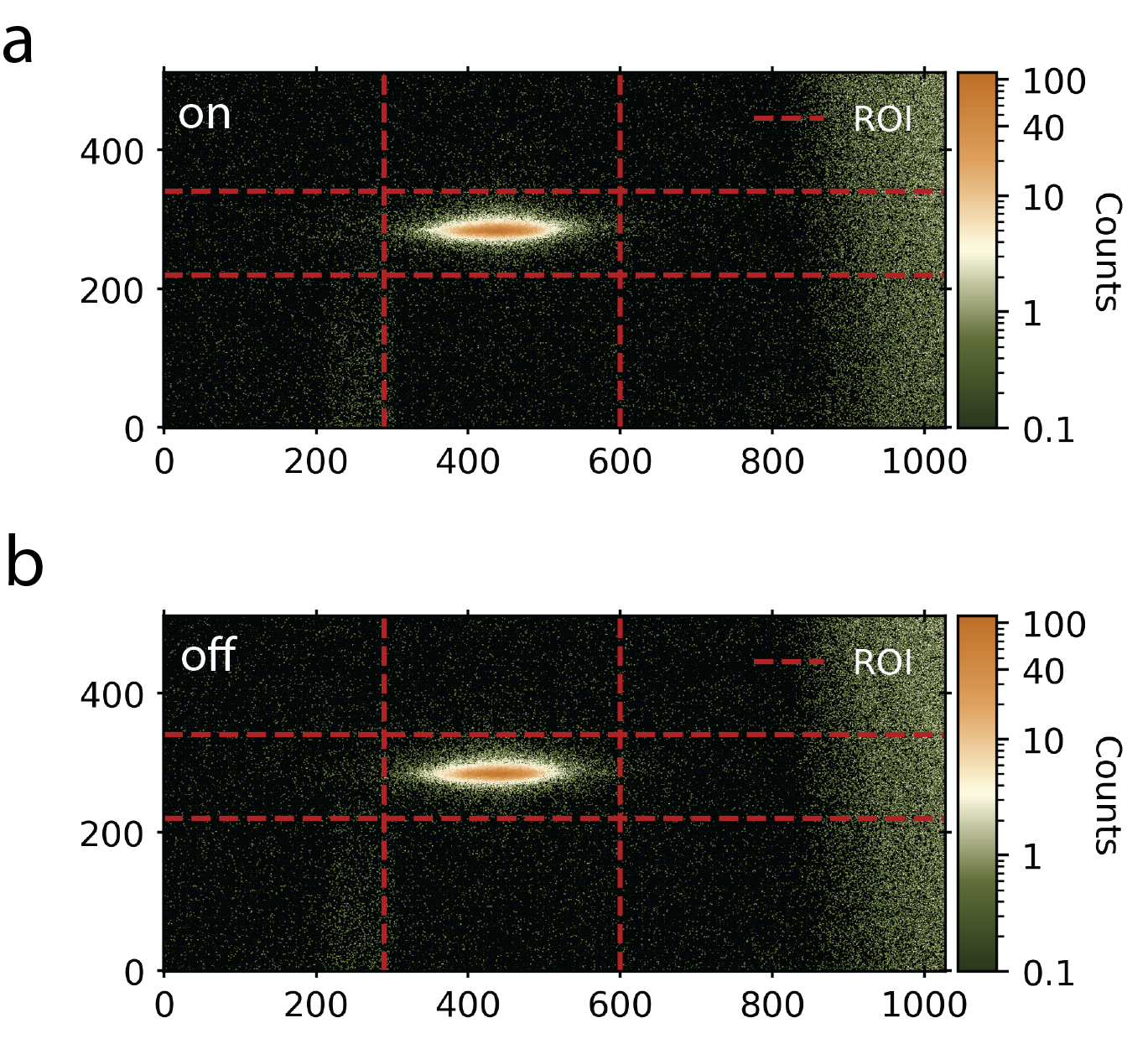}
    \caption{Two-dimensional pixel array detector image on the Eiger taken of the (002) Bragg peak \textbf{a.} after and \textbf{b.} before the arrival of the optical pump laser (at $\Delta t = 1\ \text{ns}$). Here, the peak shifts upon photoexcitation by $\sim$1 pixel upwards -- corresponding to a $2\theta$ change of $\arctan(75\ \mu\text{m}/700\ \text{mm})=6\ \text{mdeg}$.}
    \label{fig:S3}
\end{figure}

\begin{figure}[h!]
    \centering
    \includegraphics[width=0.45\linewidth]{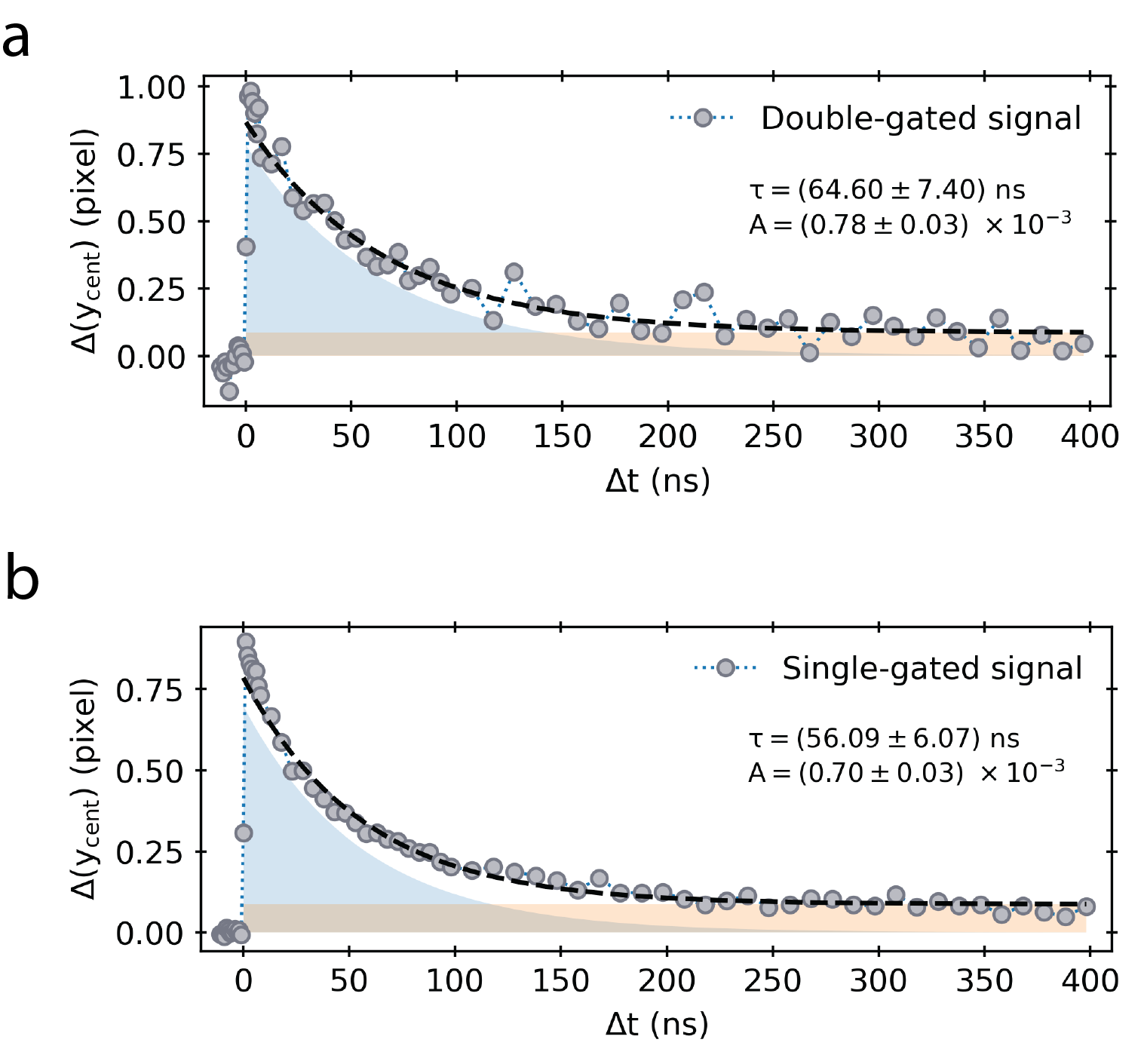}
    \caption{\textbf{a.} Signal from the double-gate mode where the $y$-centroid of the on-pulse image is subtracted from that of the off-pulse image for improved signal-to-noise and removal of gradual time effects. The on and off images correspond to those shown in Figs. 1 in the main text and \ref{fig:S3}. A fit to an exponential decay on top of a constant (blue dashed line) is shown along with the individual function contributions in shaded regions. Error values are obtained from propagating counting statistics-related error and from the fit. \textbf{b.} Similar as in \textbf{a.}, but with the single-gated mode. We note the smaller time constant compared to the double-gated mode.}
    \label{fig:S4}
\end{figure}

In our time series measurements at a constant fluence and pump-probe distance $\Delta x$, in double-gated mode, we collect two two-dimensional detector array images at each pump-probe time delay $\Delta t$: one before and one after the optical pump (Fig. \ref{fig:S3}). In single-gated mode, with higher throughput, we only take a detector snapshot after the optical pump. For the majority of the experiment, we opt for the double-gated mode. The double-gated mode, while not drastically very different from the information of single-gated mode as shown in Fig. \ref{fig:S4}, does remove a gradual linear time-dependence of the static Bragg peak in the absence of photoexcitation. The slow linear-in-time off signal tends to skew the time constant of the decaying relaxation to shorter time constants so we take care to correct this effect. Furthermore, double-gated mode removes artifacts that arise from the occasional refilling of the synchrotron with electron bunches and observed in the data as sharp spikes in intensity. More data points were taken at small $\Delta t$ to accurately capture the short time photoinduced dynamics.

\section{Intensity and width change of the signal from the pump}

\begin{figure}[h!]
    \centering
    \includegraphics[width=0.8\linewidth]{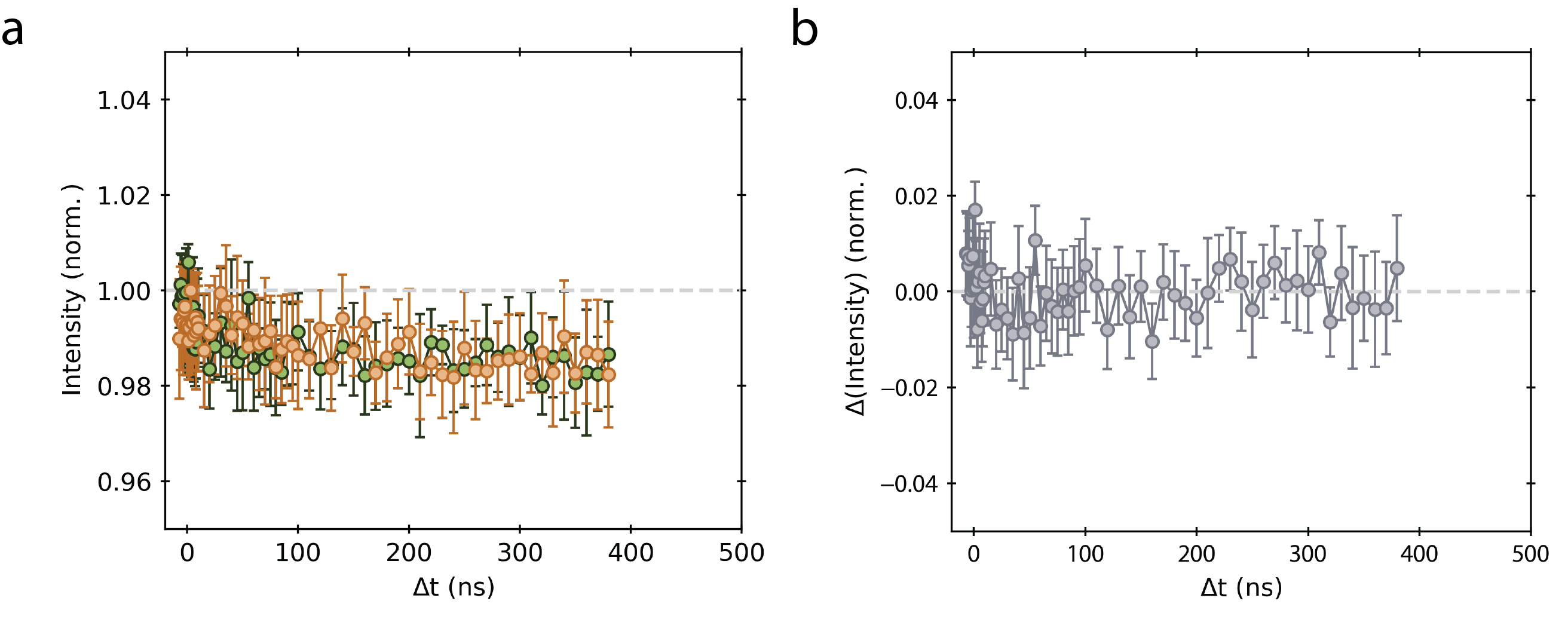}
    \caption{\textbf{a.} Integrated intensity normalized to maximum peak value as a function of pump-probe time delay $\Delta t$ when the pulse is on (green) and off (orange). \textbf{b.} Difference plot of \textbf{a.}. Error bars represent one standard deviation and originate from Poisson counting statistics.}
    \label{fig:S5}
\end{figure}

In the main text, we show the signal derived from the peak shift upon photoexcitation. Here, in this supplementary note, we show how the integrated intensity and the full width at half maximum (FWHM) of the Bragg peak on the two-dimensional array detector change with pump-probe time decay $\Delta t$ dependence. From the two-dimensional array obtained from the detector, we fit a two-dimensional Gaussian profile to the peak of the image to extract the integrated peak intensity and the peak FWHM (averaging the FWHM along the two directions and propagating the error bar). In the following plots, we plot the change in $2\theta$ angle that was converted from the $y$-centroid shift by noting the individual pixel size and the distance between the detector and the sample (as $\arctan(75(\Delta \text{pixel})\times 10^{-6}/700\times 10^{-3})$ where $\Delta \text{pixel}$ is the shift in pixels). As shown below, the integrated intensity (normalized to the maximum intensity) varies insignificantly with $\Delta t$ on top of the gradual drift described previously, only changing by a few percent within error bars (Fig. \ref{fig:S5}). 

\begin{figure}[h!]
    \centering
    \includegraphics[width=0.8\linewidth]{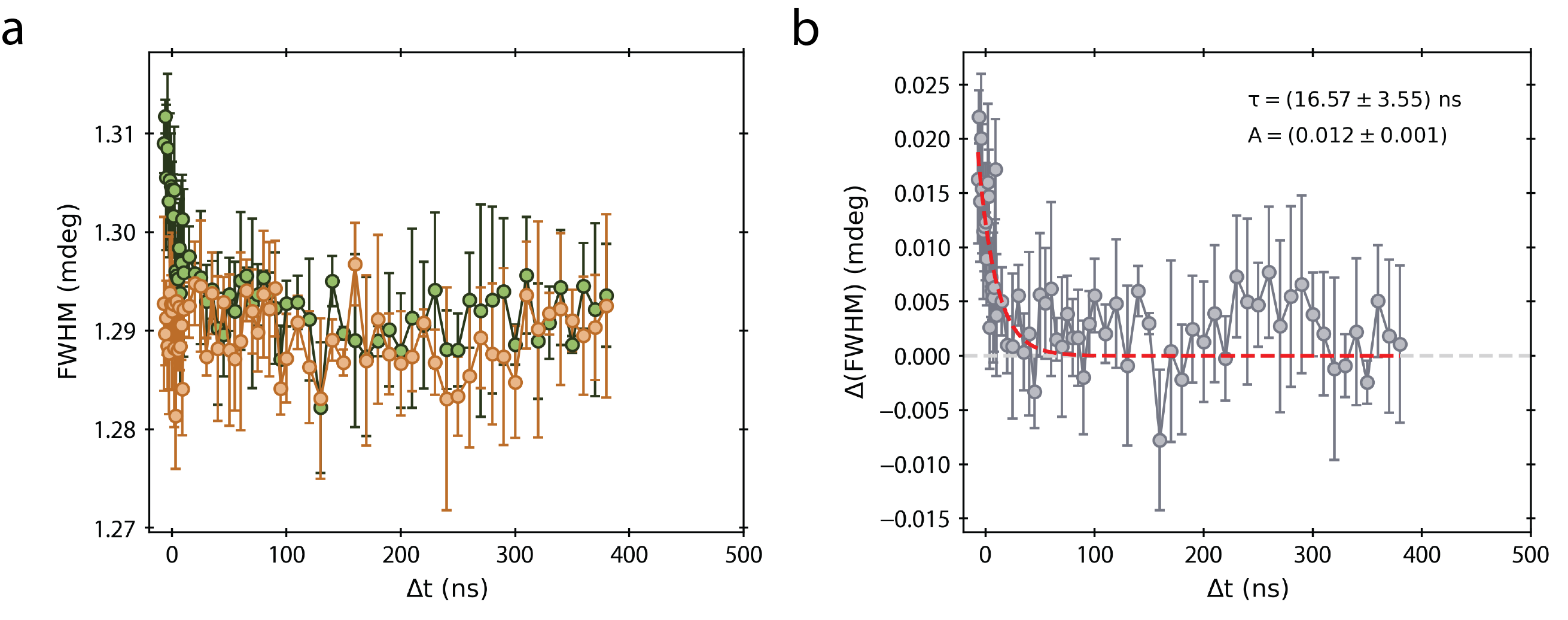}
    \caption{\textbf{a.} Full width at half maximum (FWHM) value as a function of pump-probe time delay $\Delta t$ when the pulse is on (green) and off (orange). \textbf{b.} Difference plot of \textbf{a.}. Error bars represent one standard deviation and originate from Poisson counting statistics. A exponential fit (red dashed line) is added with fitted values and their errors displayed.}
    \label{fig:S6}
\end{figure}

By contrast, the FWHM of the Bragg peak does manifest a noticeable change as a function of pump-probe time delay $\Delta t$. We show, in Fig. \ref{fig:S6}, plots of the FWHM and the difference between instances when the optical pump is on and off. The signal decays at a much shorter time constant (< 20 ns) in comparison to the $2\theta$ shift signal. We also note that this signal using the FWHM is only apparent at large enough fluence levels of the optical pump (including the fluence level we ran at for most of the experiment) and does not appear when the laser power is small. These observations suggests a generation of electronic strain or of coherent ballistic phonons at this shorter time scale compared to the longer diffusive thermal processes. The FWHM value and the change in FWHM are small ($\sim$1 mdeg and $\sim$0.02 mdeg, respectively) compared to the $2\theta$ shift (reaching 6-10 mdeg).

\section{Fitting model of time series measurements}

\begin{figure}[h!]
    \centering
    \includegraphics[width=0.7\linewidth]{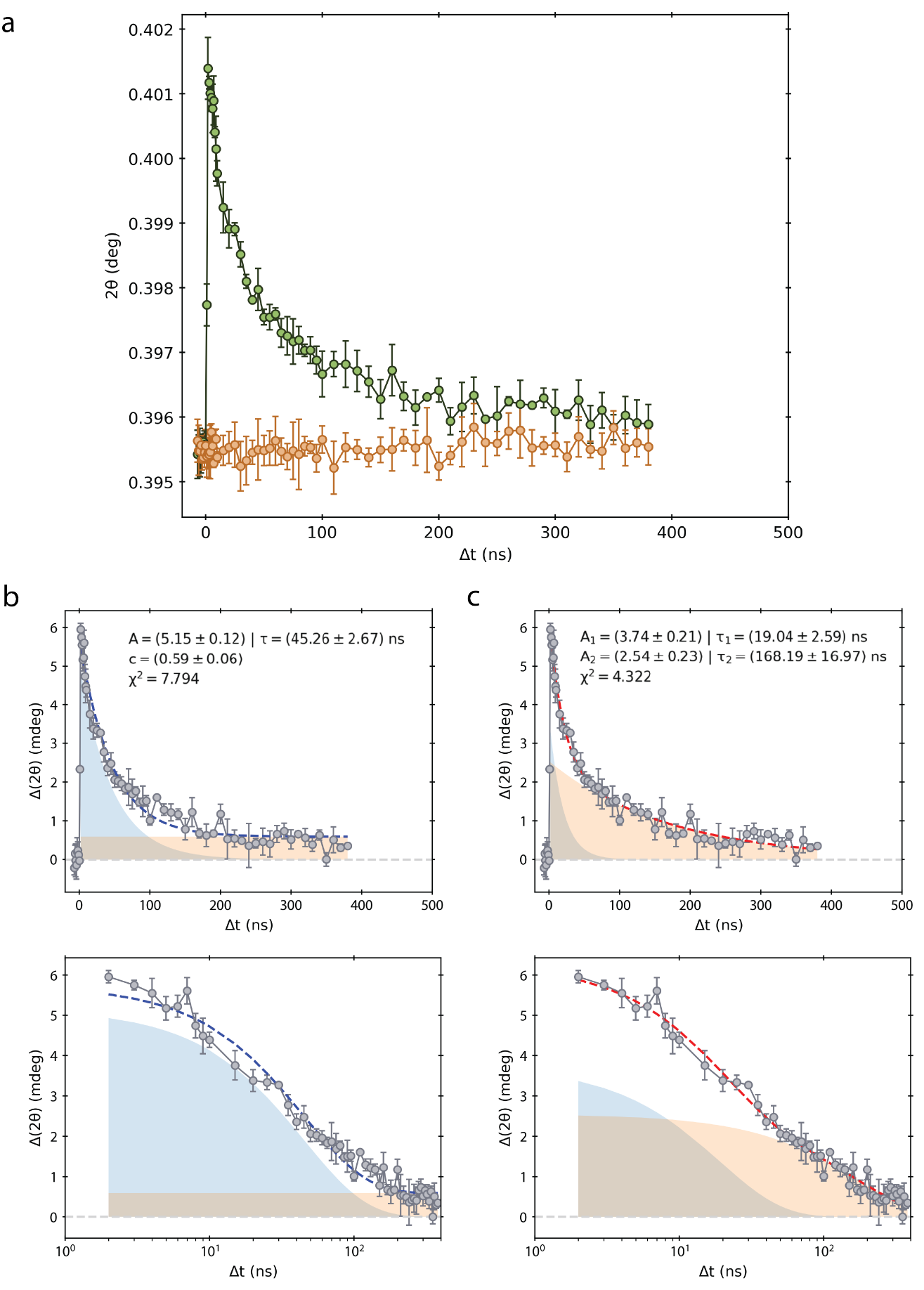}
    \caption{\textbf{a.} Plot of the $2\theta$ value of the Bragg peak as a function of pump-probe time delay $\Delta t$ with the pump on (green) and off (orange). The value of the $y$-axis corresponds to the distance away from the center of the detector and not the actual $2\theta$ value. \textbf{b.} Linear (top) and semi-logarithmic (bottom) difference plot with a fit involving an exponential added to a constant value. \textbf{c.} Linear (top) and semi-logarithmic (bottom) difference plot with a fit involving two exponential functions with a small and large time constant, respectively. In all plots, error bars represent one standard deviation that is propagated from Poisson counting statistics. Fit parameters (and their errors) and chi-square values are displayed for each corresponding fit. Contributions to the signal from each function component of the model are shown in shaded regions.}
    \label{fig:S7}
\end{figure}

As mentioned in the previous supplementary notes, it would seem that the pump-probe time delay dependence of the $2\theta$ shift would reveal two concurrent processes with distinct time scales. A hint of this hypothesis comes from the FWHM difference signal with a sub-20 ns time constant. Here, we demonstrate fits to the time series measurements using two models: 1) a double exponential with two different time constants and 2) an exponential added onto a constant value. As shown in Fig. \ref{fig:S7}, the model that includes the second exponential function improves the fit significantly (as shown by the $\chi^2$ value), especially at the low time scales seen in the semi-logarithmic plot. We interpret the signal as having a small timescale component from electronic or ballistic phonon-induced strain and a long timescale component from diffusive relaxation of the lattice. We note that the contribution to the signal from the exponential with the smaller time constant becomes larger as the fluence level increases, i.e. the $2\theta$ at early times becomes sharper at higher fluences. This indicates that it is related to a process that depends on the fluence of the optical laser -- which would be consistent with the picture of electronic strain due to more generated charge carriers. The exponential with long time constant, which we associate with the diffusive process, is characterized by $\tau >$ 60 ns. For our thermal modeling of the experiment, we use this two-component analysis to support our simplified initial temperature profile by extracting out the early-time portion of the experimental data.

\section{Fluence dependence of the optical pump laser}

\begin{figure}[h!]
    \centering
    \includegraphics[width=0.9\linewidth]{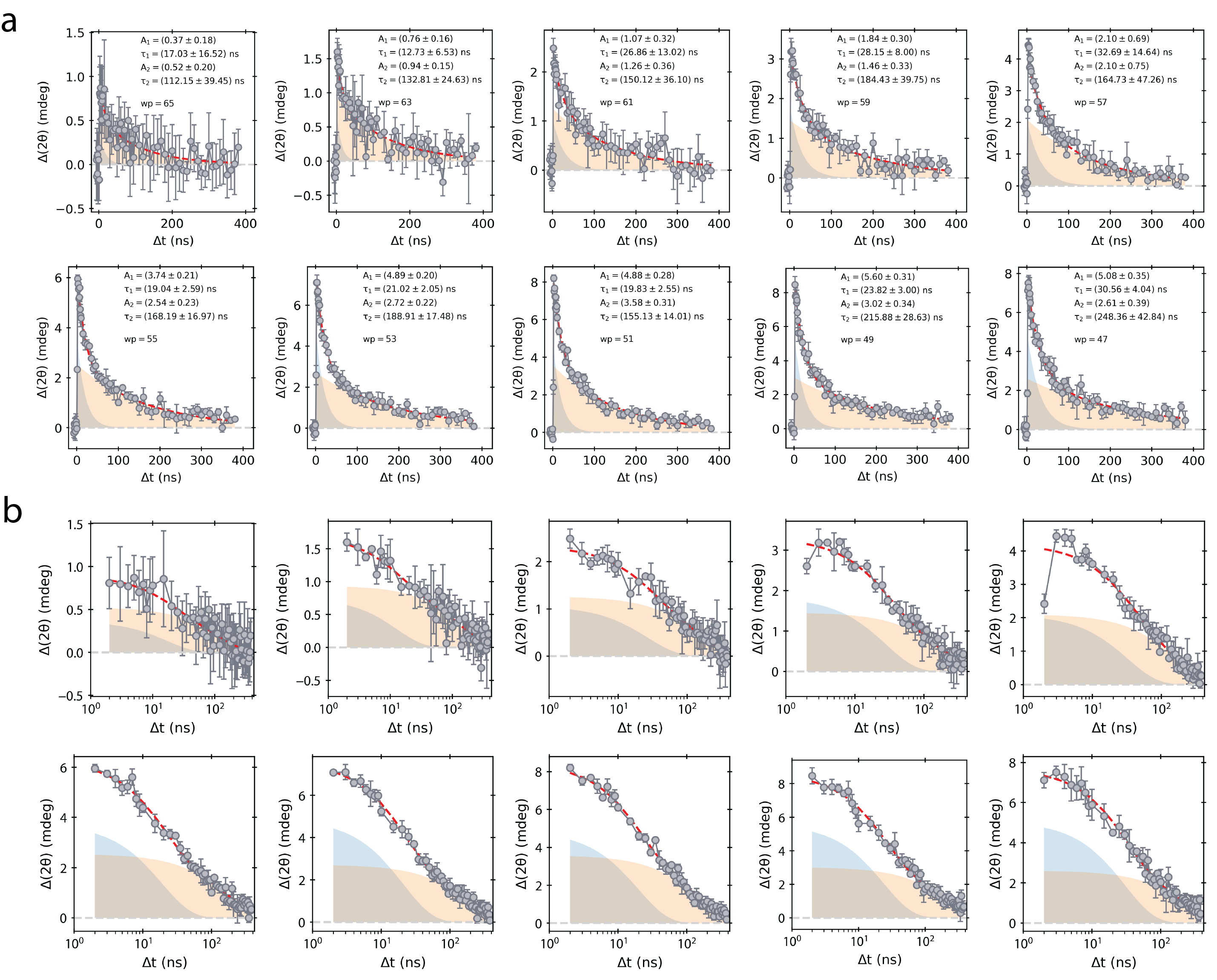}
    \caption{\textbf{a.} Plots of the $2\theta$ angular shift as a function of pump-probe time delay $\Delta t$ at different waveplate angles (and thereby, different fluence levels) from 65$^\circ$ to 47$^\circ$ and at spatial coincidence between the pump and probe ($\Delta x = 0$). \textbf{b.} Corresponding plots as in \textbf{a.}, but with the $\Delta t$ axis on a logarithmic scale. Fits are based on the double-exponential model described in previous sections. Fit parameters and their error bars are displayed. Error bars represent one standard deviation and originate from Poisson counting statistics.}
    \label{fig:S8}
\end{figure}

Fig. \ref{fig:S8} displays the fluence dependence of the optical pump on the time series measurements of the $2\theta$ angular shift. A few trends can be observed. As the fluence increases (waveplate angle decreases), the maximal $2\theta$ angular shift immediately near $\Delta t = 0$ increases monotonically up to a certain value of 53$^\circ$ or 51$^\circ$ after which the value saturates. This is due to sample degradation at high fluence which was also observe visually by eye on the sample as lines indicating sample burn. Furthermore, as the fluence increases even beyond the saturation point, the sharpness of the drop in the early time signal increases which can be observed as a larger contribution of the low-time-scale exponential function and the longer decay time for the second exponential. These observations hint at a fluence-dependent mechanisms for the early time scales such as an electronic strain. As mentioned in the main text, we chose to run the majority of the experiment at a waveplate angle of 55$^\circ$ which sits below the value at which the maximum $2\theta$ angular shift saturates.

\section{Angular shift at large pump-probe distances and laser intensity profile}

\begin{figure}[h!]
    \centering
    \includegraphics[width=0.8\linewidth]{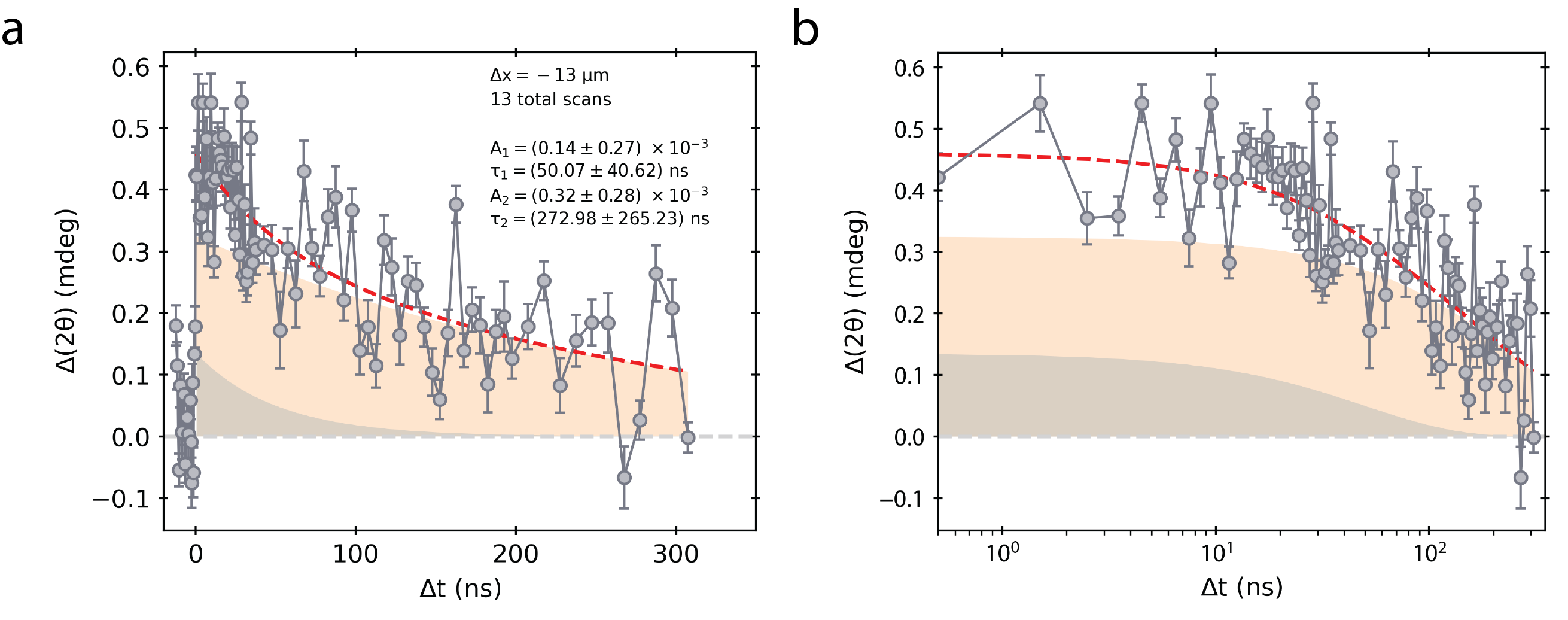}
    \caption{\textbf{a.} Plot of the $2\theta$ angular shift versus pump-probe delay time at a spatial distance of $\Delta x = 13\ \mu$m between the pump and probe from the sum of thirteen time series measurements. \textbf{b.} Corresponding plot with the $\Delta t$-axis in logarithmic scale. The red dashed line indicates a fit to a double-exponential function. Error bars indicate one standard deviation from Poisson statistics.}
    \label{fig:S9}
\end{figure}

In Fig. \ref{fig:S9}, we plot the $2\theta$ angular shift where the distance between the pump laser and the x-ray probe is 13 $\mu$m. Here, we sum over thirteen measurements of the time series to improve the statistics. The fitting error remains relatively substantial at these large pump-probe distances. The maximum $2\theta$ angular shift is 0.46 mdeg which corresponds to a lattice constant change of 6.5 fm or equivalently, a strain value of 1.25$\times 10^{-5}$.

\begin{figure}[h!]
    \centering
    \includegraphics[width=0.8\linewidth]{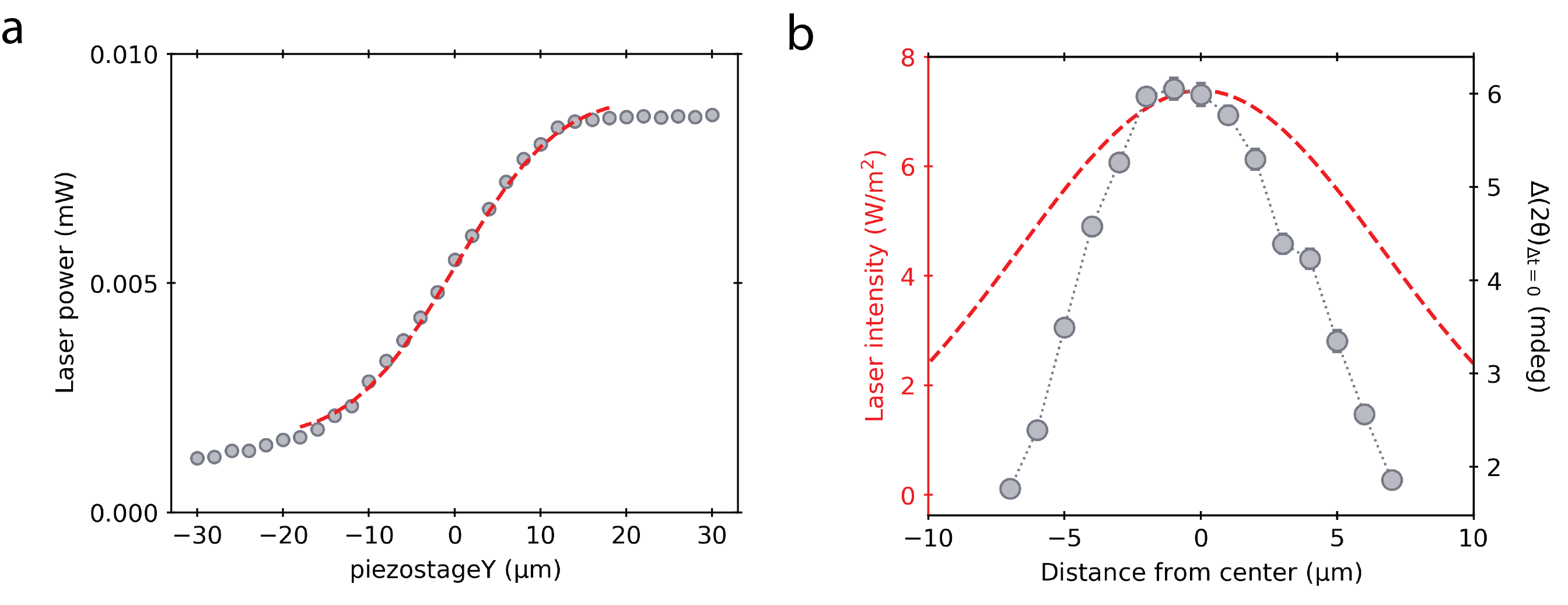}
    \caption{\textbf{a.} Knife edge scan of the beam profile for the optical pump laser. A fit is shown as a red dashed line to extract the beam width.  \textbf{b.} Re-constructed beam profile of the optical pump from the knife edge scan as a red dashed line (left axis) and maximum angular shift as a function of pump-probe distance immediately after the optical pump ($\Delta t = 0$) in grey data points (right axis). The beam profile is considerably larger than that of the angular shift.}
    \label{fig:S10}
\end{figure}

To ensure that the angular shift measured in our experiment is not purely due to the intensity profile of the optical laser, we perform a knife edge scan to measure the profile of the beam. We scan a knife edge across the face of the beam while measuring the power. The power drops as the knife edge blocks progressively more of the beam, thereby enabling us to relate the measured power as a function of the knife edge position with a beam radius. As shown on the left of Fig. \ref{fig:S10}, we plot the laser power versus the $y$ position of the piezostage controlling the laser position which displays a characteristic complementary error function form (as expected for a Gaussian-shaped beam) with
\begin{equation}
    \frac{P(z)}{P_{\text{max}}} = \frac{1}{2}\text{erfc}\left(\frac{z}{\sigma}\right)
\end{equation}
where $z$ is the position, $\sigma$ is the beam width, and erfc is the complementary error function ($1-\text{erf}$, with erf being the normal error function). When a fit is of the form above is applied to the data, a beam width of (13.3 $\pm$ 0.8) $\mu$m is extracted. We showcase the reconstructed beam intensity profile on the right of Fig. \ref{fig:S10} from the knife edge scan and plot the maximum $2\theta$ shift immediately after the optical pump for different $\Delta x$. 

While there is a slight correlation between the spatial-dependence of the angular shift with the beam intensity profile, the beam profile of the optical pump does not match that of the angular shift which has a smaller characteristic width -- thereby suggesting that other factors may explain the discrepancy. We argue that the angular shift observed in the experiment does not originate from the beam profile, but rather, most prominently, from the heat dissipation in the thin film due to this discrepancy in profile shape. Notably, it should be emphasized that in the cases where a wrinkle is involved, one can observe a clear distinction in angular shift signal between equivalent distances ($\Delta x \pm 3$ $\mu$m) depending on whether the wrinkle is there or not, which would not be explainable if the effect were purely from the beam profile.

\section{Scans of the optical pump laser positions along the orthogonal direction}

\begin{figure}[h!]
    \centering
    \includegraphics[width=\linewidth]{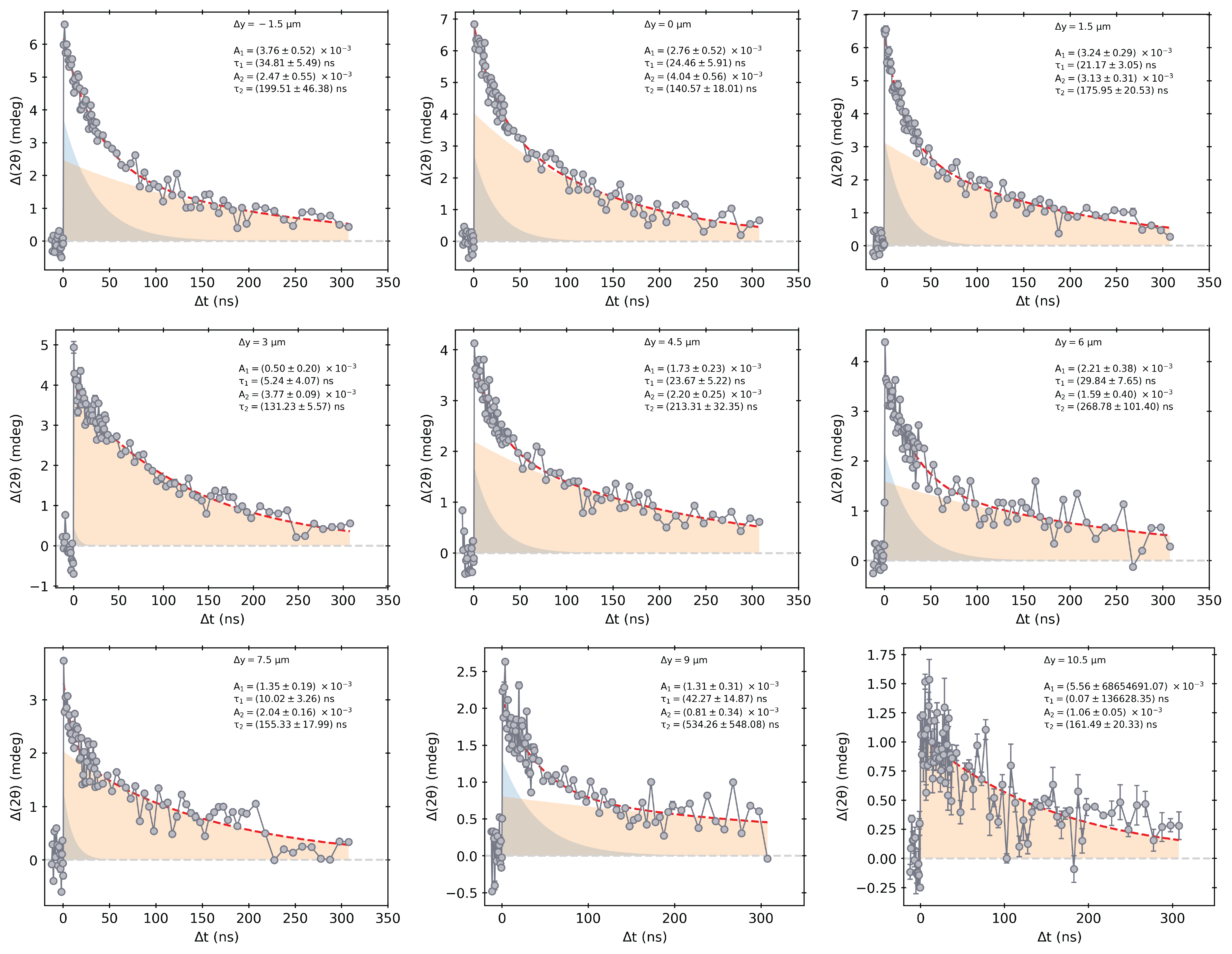}
    \caption{Plots of the $2\theta$ angular shift at different values of pump-probe distance in the vertical direction ($\Delta y$) along the $\Delta x = 0$ line on the thin film as a function of pump-probe time delay. Values of $\Delta y = -1.5\ \mu$m to $\Delta y = 10.5\ \mu$m are shown in steps of 1.5 $\mu$m. Note here $\Delta y$ is not the vector along y-axis, but the relative displacement between the pump and the probe perpendicular to the relative displacement $\Delta x$ between pump and probe. Fits are using a double-exponential function. Error bars represent one standard deviation from Poisson statistics.}
    \label{fig:S11}
\end{figure}

To verify the isotropic nature of the laser-induced heat at the center of our region of interest on the thin film, we also perform time series measurements along different positions in the orthogonal direction (forming a cross shape with the measurements in the main text). Along the $\Delta x = 0$ line, we perform measurements at 1.5 $\mu$m steps in the vertical direction at different $\Delta y$, which are shown in Fig. \ref{fig:S11}. The spatial step size is larger in the $\Delta y$ direction compared to the $\Delta x$ direction because the objective used in focusing the pump laser has a 3:2 scale between the optical pump distance and the actual distance of the laser spot on the thin film. We observe similar behavior compared to the horizontal positional scans which suggest that the heat dissipated from the laser spot is isotropic in the lateral directions.

\section{Conversion of angular shift to local temperature}

To convert the angular shift of the (002) Bragg peak of GaN into a local temperature $T_{\text{loc}}$ which will be used in our calculations involving the heat conduction equation, we use the following procedure. The initial temperature profile is directly extracted from the experiment measurements immediately after $\Delta t = 0$. To obtain the temperature profile, we first convert $\Delta(2\theta)$ into a lattice constant change $\Delta c$ using Bragg's law ($\lambda = 2d\sin(\theta)$). Here, $\lambda = 1.13$ \AA$\ $(11 keV x-ray probe) and $d$ can be obtained, for a hexagonal structure like the case of GaN, with $1/d^2 = (3/4)(h^2 + hk + k^2)/a^2 + l^2/c^2$ where $(hkl) = (002)$, $a = 3.190$ \AA, $c = 5.189$ \AA, and $2\theta = 23.8^\circ$ for our measurement. Afterwards, we can obtain the local temperature value using the linear thermal expansion coefficient $\alpha$ and relating the strain ($\Delta c/c$ or $\Delta(\theta)/\theta$) to the temperature with the formula
\begin{equation}
    \cot(\theta)\Delta\theta = -\alpha\Delta T.
\end{equation}
We set the boundary condition of the two ends of the thin film to be set at the fixed temperature of 300 K, which is the temperature of the experiment.

It is worth to mention that the out-of-plane expansion can be enhanced by a factor involving the Poisson ratio while the in-plane expansion may be forced to zero as it is epitaxially clamped to the substrate. As a result, the relation between strain and temperature should be $\Delta c/c = (1+\nu)/(1-\nu) \alpha \Delta T$ where $\nu = 0.18$ is the Poisson ratio of GaN. However, after the numerical simulation and the fitting, we notice this correction of this conversion does not affect the fitted thermal parameters. The fitted in-plane thermal conductivity ranges to $k_{\parallel}=$(93.8 $\pm$ 22.0) W/m$\cdot$K for the corrected conversion compared to the original fitted (92.8 $\pm$ 22.0) W/m$\cdot$K. The fitted localized wrinkle thermal conductivity for the corrected conversion $k_w = (20.7 \pm 0.4)$ W/m$\cdot$K compared to $k_w = (21.2 \pm 1.2)$ W/m$\cdot$K and the GaN-Si TBC value near the wrinkle defect of $G_w = (2.10 \pm 0.05)\times 10^7$ W/m$^2\cdot$K compared to $G_w = (2.12 \pm 0.05)\times 10^7$ W/m$^2\cdot$K. This small improvement may be due to the small Poisson ratio of GaN and the robustness of the whole numerical workflow. The other fitting information for the corrected conversion of the data to the temperature are also attached in the other Supplementary sections.

\section{Theoretical model}
\subsection{Three-dimensional heat transport}
In this subsection, we analyze the transport mechanism that occurs within the materials of the experiment described in this work. The size of the thin film system is on the scale of microns with a thickness of 500 nm, and the phonon mean free path is at most around 100 nm. The latter is smaller in both the in-plane and cross-plane directions. The heat diffusion characteristic time $t_{c} \approx l^2/D$ is around several nanoseconds, matching the time scale in the experiments, where $D$ is the thermal diffusivity. This can be estimated as $D=k/\rho C$ where $k > 0$ is the thermal conductivity, $\rho$ is the mass density, and $C$ is the specific heat capacity.

As a result, the thermal transport mechanism should be diffusion-dominated and can be described by Fourier's law and the heat conduction equation \cite{chen2005}. The general heat conduction equation is
\begin{equation}
C \rho \frac{\partial T(\mathbf{r}, t)}{\partial t} = \nabla \cdot (\overleftrightarrow{\mathbf{k}} \nabla T(\mathbf{r}, t)),
\end{equation}
where $T(\mathbf{r},t)$ is the local effective temperature with dependencies on position $\mathbf{r}$ and time $t$, $C$ is the heat capacity, and $\rho$ is the mass density. This equation governs the transport behavior for $\mathbf{r} \in \Omega$ where $\Omega$ is the area within the material. Here, $\overleftrightarrow{\mathbf{k}}$ is typically a thermal conductivity tensor, but we take a simplified case by assuming that the non-diagonal part vanishes. In this case, only the diagonal elements $k_{xx}=k_{yy} = k_{\parallel}$, and $k_{zz} = k_{\perp}$ are nonzero in which $\overleftrightarrow{\mathbf{k}}$ has spatial dependence.

$\mathbf{r} \in \partial \Omega$ represents the boundary or, in our case, the junction or contact between different materials. For the experiment described here, $\partial \Omega$ is the contact area between the GaN thin film and the Si substrate. We discuss how to solve for this boundary in the following subsection using a thermal boundary conductance (TBC) simulation.

\subsection{Boundary condition: Simulate TBC using the heat conduction equation}
In general, we should consider the TBC along the cross-plane direction at the boundaries between different materials. For our material system, we only need to consider the main interface between GaN and Si, which otherwise can be further generalized in principle into a system with multiple interfaces. TBC is defined as the thermal flux per unit area per temperature difference. This can be expressed through the boundary condition
\begin{equation}
\begin{split}
        -k_{zz,\mathrm{GaN}}\frac{\partial T_{\rm GaN,b}}{\partial {z^{+}} } = G_{\rm{GaN,Si}} \left(T_{\rm{Si,b}}-T_{\rm{GaN,b}}\right),\\-k_{zz,\mathrm{Si}}\frac{\partial T_{\rm Si,b}}{\partial {z^{-}} } = G_{\rm{GaN,Si}} \left(T_{\rm{GaN,b}}-T_{\rm{Si,b}}\right).
\end{split}
\end{equation}
$G_{\rm{GaN,Si}}$ is the TBC of the interface between Si and GaN. A subscript $b$ represents the spatial location on the contacting boundary layer between the GaN thin film and the Si substrate. It indicates the position at the bottom of GaN and the top of the Si substrate Si. $T_{\rm Si,b}$ and $T_{\rm{GaN,b}}$ are the temperature profiles at the contacting boundary layer for Si and GaN, respectively. As a result, this boundary condition provides a normal vector of the temperature gradient along the out-of-plane direction.

\subsection{Initial condition: Simplified initial profile}
In these calculations, we set the mass density of GaN as $\rho = 6.08$ g/cm$^3$ and the heat capacity as 35.4 J/(mol$\cdot$K). The characteristic thermal redistribution time along the cross-plane direction for the thin GaN film is estimated to be approximately 8 ns if we assume that the cross-plane thermal conductivity is 80 W/(m$\cdot$K) which is close to the upper bound measured from our TDTR experiments. This implies that the film requires this period to reach thermal equilibrium along this direction.

To simplify the problem, we neglect the first couple of nanoseconds which allows the temperature profile within the thin film to fully redistribute. Both the FWHM measurement and other experimental observations suggest the presence of electronic strain effects which may potentially arise from complex electron-photon interactions that occur on shorter timescales than thermal diffusion. To ensure a more accurate comparison between the simulation and the experimental data, the introduction of a delayed initial time is reasonable to account for these strain effects. This approach minimizes the influence of the latter provided that the delay is not excessive as cross-plane heart transport between the thin film and the substrate remains minimal, while ensuring that the initial temperature profile remains accurately represented. Consequently, we can model the cross-plane transport as a two-timescale problem where the thermal redistribution within the thin film occurs rapidly on a small timescale whereas the heat conduction along the substrate and the heat transfer between the thin film and the substrate are comparatively slower on longer timescales. If we neglect the rapid initial phase and focus only on the slower dynamics, we can simplify the problem by adopting a coarser spatial mesh (on the order of microns like our experiment) and redefining the initial time.

In this case, the substrate can be reasonably assumed to remain at room temperature as the early-stage heat transfer from the thin film to the substrate is minimal and the TBC is relatively low. One must take note that this approach may lead to a slight underestimation of thermal parameters as it involves a necessary trade-off in accuracy for the sake of computational efficiency.

\section{Theoretical model for wrinkle-affected thermal transport}
We describe how we modify the thermal model to include the effect caused by the presence of a wrinkle on the thin film. To simplify the discussion without loss of generality, we consider the situation in one dimension along the in-plane direction (such as along the x direction) and then generalize to the three-dimensional case later on. We consider that the presence of the wrinkle would affect the thermal conductivity $k_{\parallel}$ locally, leading to spatial inhomogeneity of the thermal conductivity. As a result, we rewrite spatially dependent $k_{\parallel} = k_{\parallel,\text{hom}} + Q(x,x_w,\epsilon) \Delta k_w$ where $x_w$ is the position of the wrinkle, $k_{\parallel, \text{hom}}$ is the homogeneous thermal conductivity, $\Delta k_w$ is the wrinkle-induced change of the thermal conductivity, and $Q(x,x_w,\epsilon) = \exp\left(-\frac{(x-x_w)^2}{2\epsilon^2}\right)$ is the Gaussian kernel function with positional parameter $x_w$ and width parameter $\epsilon$. The latter is a small parameter used to control the global and local coupling. When $x = x_w$, we define $k_w = k_{\parallel,\text{hom}} + \Delta k_w$. This treatment can be considered as a multiscale modeling to resolve singularities as shown in Ref. \cite{weinan2011}. After inserting the wrinkle-induced term back into the original equation, we obtain
\begin{align}
\begin{split}
    C\rho\frac{\partial T}{\partial t} &= k_{\parallel}\frac{\partial^2 T}{\partial x^2} + \frac{\partial k_{\parallel}}{\partial x}\frac{\partial T}{\partial x}\\
    &= k_{\parallel,\text{hom}}\frac{\partial^2 T}{\partial x^2} + Q(x_w, \epsilon)\left(\Delta k_w\frac{\partial^2 T}{\partial x^2} - \Delta k_w\frac{2(x-x_w)}{2\epsilon^2}\frac{\partial T}{\partial x}\right)\\
    &\approx k_{\parallel,\text{hom}}\frac{\partial^2 T}{\partial x^2} + \delta_{x, x_w}W,
\end{split}
\end{align}
where $W = \Delta k_w(\partial^2 T/\partial x^2) - \Delta k_w((2(x-x_w))/(2\epsilon^2))(\partial T/\partial x)$ can be considered as a source term, which contains dependencies on $\Delta k_w(\partial T/\partial x)$ and $\partial^2 T/\partial x^2$. If we assume the parameter $\epsilon$ is small, then the Gaussian kernel function can be approximated as $\delta_{x, x_w}$. In the numerical calculation presented later on, as space is discretized, we can linearly approximate the kernel with a trapezoid with $Q\approx 1-|x-x_w|/(2\epsilon)$ if $|x-x_w|\leq 1$ and $Q = 0$ otherwise. We can attribute and model the effect of the wrinkle as a temperature profile-dependent point-like source or sink. The temperature profile-dependent heat source/sink induced by the presence of the wrinkle may lead to non-reciprocal thermal transport. The latter was inferred from asymmetrical measurements in our experiment and confirmed through our numerical calculations. By generalizing to the three-dimensional case, we have
\begin{align}
\begin{split}
    C\rho\frac{\partial T}{\partial t} &= k_{xx}\frac{\partial^2 T}{\partial x^2} + \frac{\partial k_{xx}}{\partial x}\frac{\partial T}{\partial x} + k_{yy}\frac{\partial^2 T}{\partial y^2} + k_{zz}\frac{\partial^2 T}{\partial z^2}\approx \nabla \cdot
(\overleftrightarrow{\mathbf{k}}_{\rm{hom}} \nabla T) + \delta_{x, x_w}W
\end{split}
\end{align}
where we have assumed the defect-cased heterogeneity only arises along the $x$-axis and that the defect acts as a heat source or sink in the three-dimensional case, here $\overleftrightarrow{\mathbf{k}}_{\rm{hom}}$ is the homogeneous thermal conductivity tensor.

We briefly explain with a microscopic picture. Without loss of generality, we discuss the situation with a one-dimensional picture (along the x axis) with general thermal conductivity $k$, homogeneous thermal conductivity $k_{\text{hom}}$, the change of thermal conductivity caused by wrinkle $\Delta k_{w}$, the thermal conductivity $k_{w}$ at the wrinkle, while a similar one can also apply in three dimensions. The thermal conductivity can be characterized through the Callaway model with the isotropic and single-mode approximation as 
\begin{equation}
    k = \frac{1}{3}\int_0^{\omega_{\text{max}}}C_s(\omega)\tau(\omega)v_g^2(\omega)\ d\omega
\end{equation}
where $C_s$ is the spectral heat capacity, $\tau$ is the relaxation time, and $v_g$ is the phonon group velocity. The wrinkle emerges as a spatial-dependent contribution for this term. We consider that the region close to the wrinkle can be described with a wrinkle-induced phonon scattering rate ($\tau_{pw}^{-1}$), phonon-phonon scattering rate ($\tau_{pp}^{-1}$), and phonon-impurity scattering rate ($\tau_{pd}^{-1}$) such that $\tau_w^{-1} = \tau_{pw}^{-1} + \tau_{pp}^{-1} + \tau_{pd}^{-1}$. In the non-singularity region, we only have the phonon-phonon scattering rate and the phonon-impurity scattering rate such that the corresponding homogeneous relaxation rate is $\tau_{\text{hom}}^{-1} = \tau_{pp}^{-1} + \tau_{pd}^{-1}$.

The corresponding expressions for the thermal conductivity are given by
\begin{align}
\begin{split}
    k_{\text{hom}} &= \frac{1}{3}\int_0^{w_{\text{max}}} C_s(\omega)\tau_{\text{hom}}(\omega)v_g^2(\omega)\ d\omega\\
    \Delta k_w &= \frac{1}{3}\int_0^{\omega_{\text{max}}}(\tau_w(\omega) - \tau_{\text{hom}}(\omega))v_g^2(\omega)\ d\omega\\
    k_w &= \frac{1}{3}\int_0^{w_{\text{max}}} C_s(\omega)\tau_w(\omega)v_g^2(\omega)\ d\omega
\end{split}
\end{align}
where $k_w$ designates the thermal conductivity at the exact position of the wrinkle. The total thermal conductivity can be rewritten as
\begin{equation}
    k = k_{\text{hom}} + Q(x_w,\epsilon)\Delta k_w=\frac{1}{3}\int_0^{\omega_{\text{max}}} C_s(\omega)(\tau_{\text{hom}}+Q(x_w,\epsilon)(\tau_w(\omega)-\tau_{\text{hom}}))v_g^2(\omega)\ d\omega.
\end{equation}

The defect region can be considered as a perturbation of the original thermal conductivity, coupling with both the singularity and the homogeneous region. The spatial coupling can be effectively achieved by adding the function $Q(x_w,\epsilon)$ as the strength coefficient. We can assume that the wrinkle-induced potential acts as a two-level model that modifies the phonon scattering. The higher energy level represents the thermal excitation of the wrinkle. It is reasonable to assume that the higher energy level would be relieved or degenerated with decreasing distance between the wrinkle, as this would make the wrinkle easier to excite and create more resistance. The excited energy is stored inside the wrinkle through elastic deformation. This explains the singularity-like feature of this term in the equation. Furthermore, it provides a satisfactory picture of the temperature profile dependence in the effective heat conduction equation, as it leads to a possible non-reciprocal transport. The large thermal gradient characterized by the local effective temperature gradient leads to higher excitations via the $\partial T(x,t)/\partial x\cdot\partial k/\partial x$ term where $\partial k/\partial x$ can be understood as the gap between the two levels and $\partial T(x,t)/\partial x$ represents the different excitations. The asymmetric position of the wrinkle introduces inversion-symmetry breaking which enables the non-reciprocal transport.

Wrinkles also significantly impact the TBC between the thin film and the substrate. While a comprehensive theoretical framework describing the influence of defects, including wrinkles, on TBC is lacking, we qualitatively outline the potential mechanisms behind these effects. Due to their characteristic size being in the micron scale, wrinkles induce residual strain fields extending over large areas such that strain effects typically propagate over long ranges in elastic media. These residual strain fields disrupt the uniformity of the interface affecting both the mechanical contact and the vibrational coherence between the thin film and the substrate. There are two main ideas behind this disruption. First, the wrinkle may reduce the effective contact area between the two materials which creates localized thermal resistances that collectively lower the overall TBC. Second, the residual strain may alter the atomic-scale interactions at the interface which weakens the phonon coupling and introduces additional scattering sites thereby hindering phonon transmission across the interface. Furthermore, the strain field spreading from the wrinkle may cause spatial variations in the elastic properties of the interface. These variations could exacerbate phonon mismatches and scattering and lead to a reduction in the efficiency of heat transfer. As a result, the presence of a wrinkle not only affects the immediate vicinity of the defect, but can also degrade the sensitive TBC across the entire detected region.

\section{Numerical simulation of the thermal model}
As discussed in a previous section, we neglect the experiment data contained in the first $\sim$8 ns of the experiment and wait for after the thermal redistribution within the thin film along the cross-plane direction. We extract the corresponding temperature profile at different spatial positions after the initial redistribution period to construct the discretized initial temperature profile with the experimental noise. Afterwards, we fit and smooth the temperature profile with an exponential function $A \cdot e^{\frac{x-b}{\tau}}$ to simulate the thermal pump where $A$, $b$, $\tau$ are the fitting parameters. For the homogeneous transport without wrinkles, we assume $b = 0$. The temperature profile with the wrinkle is asymmetrical due to the presence of the wrinkle acting as a heat source or sink. To treat the initial temperature profile of this scenario, we separate the profile at the maximal point and fit the separated temperature profile correspondingly with the exponential function to reflect the asymmetrical feature.

We simulate the results by numerically solving the governing equations of the previous sections with the finite difference method. The wrinkle-induced thermal conductivity is simulated through a triangle-like function with the spike sitting at the wrinkle position. To simplify the calculation, the thermal conductivity is considered homogeneous on each grid point except for those that correspond to positions around the wrinkle. We attempt to solve the equations with the forward Euler scheme of the finite difference method while simultaneously accounting for both the wrinkle and the special boundary condition caused by the TBC. We discretize using
\begin{equation}
\begin{split}
    \delta T_{2,\mathbf{s}}(\mathbf{r},i) &= T(\mathbf{r}+\mathbf{s},i) - T(\mathbf{r},i)\\
    \delta T_{1,\mathbf{s}}(\mathbf{r},i) &= T(\mathbf{r},i) - T(\mathbf{r}-\mathbf{s},i)
\end{split}
\end{equation}
where $T(\mathbf{r},i)$ represents the local temperature profile at position $\mathbf{r}$, time $i$, and $\mathbf{s}$ represents the unit vector of the axis $x$, $y$, $z$. The external boundary condition is imposed through $T(\mathrm{b_{s,e}},i) = 300$ K ($\mathrm{b_{s,e}}$ indicates the position at the boundary between the system and the external environment). The spatial derivatives are approximated as
\begin{equation}
\begin{split}
    \frac{\partial T(\mathbf{r},i)}{\partial s} \approx \frac{\delta T_{2,\mathbf{s}}(\mathbf{r},i) + \delta T_{1,\mathbf{s}}(\mathbf{r},i)}{2\delta s}\\
    \frac{\partial ^2 T(\mathbf{r},i)}{\partial x^2} \approx \frac{\delta T_{2,\mathbf{s}}(\mathbf{r},i) - \delta T_{1,\mathbf{s}}(\mathbf{r},i)}{(\delta s) ^2}
\end{split}
\end{equation}
where $s$ represents the different components $x$, $y$, $z$. $\delta s$ is the corresponding grid spacing.
The time derivative is approximated as 
\begin{equation}
\begin{split}
    \frac{\partial T(\mathbf{r},i)}{\partial t} \approx \frac{T(\mathbf{r},i+1)-T(\mathbf{r},i)}{\delta t}
\end{split}
\end{equation}
where $\delta t$ is the step size in the time domain.
The forward Euler scheme is expressed as
\begin{equation}
\begin{split}
    T(\mathbf{r},i+1) = T(\mathbf{r},i) + \frac{\delta t }{C\rho}\sum_{s}\frac{k_{ss}(T(\mathbf{r}+\mathbf{s},i)-2T(\mathbf{r},i)+T(\mathbf{r}-\mathbf{s},i))}{(\delta s)^2} \\+ \frac{\delta t }{C\rho}\frac{k_{xx}(\mathbf{r}+\hat{\mathbf{x}})-k_{xx}({\mathbf{r}-\hat{\mathbf{x}}})}{2\delta x}\frac{T(\mathbf{r}+\hat{\mathbf{x}},i)-T(\mathbf{r}-\hat{\mathbf{x}},i)}{2\delta x}.
\end{split}
\end{equation}
Notice that we only assume the wrinkle affecting the thermal conductivity along the x direction. $\hat{\mathbf{x}}$ is the unit vector along the x-axis. The mass density and the heat capacity of GaN are taken to be 6.08 g/cm$^{3}$ and 35.4 J/mol K at room temperature \cite{ravindra2017}, respectively. The spatial mesh spacing $\delta x = \delta y = \delta z$ is taken to be 1 $\mu$m to correspond with the experimental setup and the temporal mesh spacing $\delta t$ is 0.5 ns. We set this large spatial mesh spacing to simplify the simulation by ignoring the detailed temperature profiles reliant on the thin film thickness dependence. This is equivalent to assuming that the temperature distribution along the out-of-plane (z) direction of the thin film is uniform, which is reasonable after we reset the initial simulation time after several ns to allow the z-direction thermal redistribution to reach a uniform distribution. For the homogeneous region, the stability of the forward Euler scheme requires $\delta t \leq ((\delta x)^2 C\rho) / (2 k_{xx} + 2 k_{yy}+ 2k_{zz})$ from von Neumann stability analysis. Based on our numerical setup and assuming $k_{xx} = k_{yy}= k_{zz}$, the highest-tolerated thermal conductivity for the forward Euler scheme is calculated to be 856.67 W/m$\cdot$K, much larger than our situation. We will take the calculated thermal conductivity bound of the forward Euler scheme as the approximate algorithmic stability bound.

We discuss how to implement the TBC calculation. We add two ``virtual layers'' between the two interface layers: one belongs to the GaN thin film and the other one belongs to the Si substrates. From top to bottom, we have the first computing layer representing the thin film GaN which grows on the substrate. Following this layer is a virtual one added for calculating the heat flux from the GaN thin film to the substrate, followed by another virtual layer for the Si substrate. Lastly, there is the top layer of the Si substrate representing the real temperature profile at that location. The temperature difference at the boundary should be the corresponding one between the top layer representing GaN thin film and the overall fourth layer,r representing the top layer of the Si substrate. The two virtual layers are only present for calculating the directional temperature gradient conveniently and are not reflective of real temperature profiles. To discretize the boundary condition, we have
\begin{equation}
\begin{split}
        -k_{zz,\mathrm{GaN}}\frac{ T_{\rm GaN,b}(i+1) - T_{\rm GaN,v}(i+1)}{\delta z } &= G_{\rm{GaN,Si}} (T_{\rm{Si,b}}(i+1)-T_{\rm{GaN,b}}(i+1)),\\-k_{zz,\mathrm{Si}}\frac{T_{\rm Si,b}(i+1)-T_{{\rm Si,v}}(i+1)}{\delta z  } &= G_{\rm{GaN,Si}} (T_{\rm{GaN,b}}(i+1)-T_{\rm{Si,b}}(i+1))
\end{split}
\end{equation}
where $T_{\rm Si,b}$ and $T_{\rm GaN,b}$ are calculated from the numerical scheme of the heat conduction equation mentioned above, $T_{\rm Si,v}$ and $T_{\rm GaN,v}$ are the temperature profiles at the virtual layer. Note that when calculating the Laplacian of the actual boundary layer, we take the temperature profile of the virtual layer at the time step $i$ to help determine the Laplacian through the gradient. As a result, we only need to calculate the temperature profile for the virtual layer from the boundary condition, which is equivalent to solving a linear system above. This determines the temperature on the virtual layer and fixes the heat flux to keep this boundary condition at each time step. This virtual layer can also apply to the boundary between the system and the external environment by assuming thermal insulation. This is equivalent to having the corresponding TBC equal to 0 and fixing the environment temperature as $T_{\rm env} = 300$ K. 

To determine the thermal conductivity from the experimental data, we design the evaluation loss function to characterize the deviation between the simulated spatial and time-resolved temperature profiles and the experimental data. By solving the optimization problem of this loss function, we determine the best-fitted thermal conductivity parameters. We use the residue as well as the first spatial and time derivatives of the temperature profile between the theoretically calculated results and the experimental data. We also add the correction term to reduce the deviation between the experimental data and the simulation of the largest temperature difference within the temperature profile at a fixed time, which represents how quickly the overall heat dissipates in the system. Here, we choose the pump-probe direction along the x-axis to simplify the comparison between experimental data and the modeling temperature profile. The comparison then reduces to the case of one spatial dimension along $x_i$ and one temporal dimension along $t_j$. The loss function can be written as
\begin{equation}
\begin{split}
    \mathcal{L} = &\sum_{x_i,t_j}\left|T_{\mathrm{m}}(x_i,t_j)-T_{\mathrm{s}}(x_i,t_j)\right|+\sum_{x_i,t_j}\left|\frac{\partial T_{\mathrm{m}}(x_i,t_j)}{\partial x_i}-\frac{\partial T_{\mathrm{s}}(x_i,t_j)}{\partial x_i}\right|+\sum_{x_i,t_j}\left|\frac{\partial T_{\mathrm{m}}(x_i,t_j)}{\partial t_j}-\frac{\partial T_{\mathrm{s}}(x_i,t_j)}{\partial t_j}\right|\\& + \sum_{t_j}\left|\left(\max\limits_{x_i}T_{\mathrm{m}}(x_i,t_j)-\min\limits_{x_i}T_{\mathrm{m}}(x_i,t_j)\right)-\left(\max\limits_{x_i}T_{\mathrm{s}}(x_i,t_j)-\min\limits_{x_i}T_{\mathrm{s}}(x_i,t_j)\right)\right|
\end{split}
\end{equation}
where $T_{\mathrm{m}}$ and $T_{\mathrm{s}}$ are the experimentally measured and calculated temperature profiles, respectively.

To determine the in-plane thermal conductivity of the GaN film, we first fix the TBC and cross-plane thermal conductivity of the film extracted from the TDTR experiments as $k_{\perp}=$ 65 W/m$\cdot$K and 2.82$\times$10$^7$ W/m$^2\cdot$K. We use the experimental data without the wrinkle and calculate the loss function over the whole range to find the minimized value. The minimization algorithm can be a gradient-based method such as the limited-memory Broyden–Fletcher–Goldfarb–Shanno algorithm, which can be used directly from the Python-based scipy.optimize package.

However, this loss function does not apply appropriately for the wrinkle case as it will lead to highly suppressed in-plane thermal conductivity or TBC which is inconsistent between the experimental temperature profile and the simulated one. To correct this, we impose another condition which is to take the curvature of the temperature profile around the wrinkle position into account. This means we add the correction factor for the curvature around the wrinkle site defined as
\begin{equation}
    \mathcal{L}_{\rm c} =  \left|\frac{\partial ^2T_{\mathrm{m}}}{\partial x^{2}}- \frac{\partial ^2T_{\mathrm{s}}}{\partial x^{2}}\right|.
\end{equation}

Only considering the experiments related to the wrinkle will cause overfitting: the fitted results will diminish TBC or the in-plane conductivity around the wrinkle part individually. This fit does not reflect the features observed from the experimental data, which are notably, the asymmetric transport caused by the change of the in-plane thermal conductivity locally around the wrinkle and the total reduction of the thermal dissipation. To address this issue of overfitting, we add regularization to the loss function and assume the change of the TBC is relatively small due to the possible residual strain field. We add the loss function for the experimental data without wrinkle to fit TBC and the in-plane conductivity around the wrinkle simultaneously. The loss function becomes
\begin{equation}
    \mathcal{L}_{\rm tot}(G_{\mathrm{GaN,Si}}, k_{w}) =  \mathcal{L}_{\mathrm{non-wrinkle}}(G_{\mathrm{GaN,Si}}, k_{w}) + \mathcal{L}_{\mathrm{wrinkle}}(G_{\mathrm{GaN,Si}}, k_{w}).
\end{equation}

We require the change of the TBC due to the wrinkle to be relatively small to avoid overfitting and the TBC is regularized by the case without a wrinkle. To characterize the contribution of the wrinkle, we fix the thermal conductivity determined from the experiment without the wrinkle and allow the local change of the thermal conductivity and the TBC. We determine the sets of values that minimizes the loss function after scanning through the entire parameter space for these parameters. This loss function may have a very complex landscape such that the gradient-based optimization algorithm may not be proper for this situation. In that case, a differential evolution method, which is a meta-heuristic global optimization algorithm, may find superior usage. To ensure that the algorithm converges to the global minimum, we run the algorithm ten times to determine the final parameters. The error bar for these parameters is obtained by fixing the other two parameters through the error bound obtained from experimental data.

To validate the necessity of fitting both the TBC and the local thermal conductivity around the wrinkle simultaneously, we performed optimizations by varying each parameter individually. When fitting both parameters simultaneously, the optimization from wrinkle-related data yields residual errors of 65.05 and 25.27, respectively. In contrast, fitting only the local thermal conductivity around the wrinkle using the same data results in a fitted value of $k_w =  9.00$ W/m$\cdot$K with higher corresponding residual errors of 66.02 and 26.39. For the case of fitting only the TBC, the optimized TBC value is $9.45 \times 10^{6}$ W/m$^2\cdot$K with an even higher residual error of 70.52. These results demonstrate that simultaneous fitting of both TBC and $k_w$ provides an improved match to the experimental data, all while underscoring the importance of considering the impact of the wrinkle on the TBC. For the case of the corrected conversion of the data to the local temperature, the residual errors are 45.20 and 17.56 for the simultaneous fit, while it's 45.88 and 18.32 for the single fit of the local thermal conductivity around the wrinkle. For the case of ftting only the TBC, the optimized TBC value is $9.48 \times 10^{6}$ W/m$^2\cdot$K with an even higher residual error of 48.98. These results align well with the original case, while the residual errors are reduced perhaps due to the improved temperature calibration but not affect the final fitted results a lot.

\section{Monte Carlo fitting algorithm results of FDTR measurements}
\begin{figure}[h!]
    \centering
    \includegraphics[width=\linewidth]{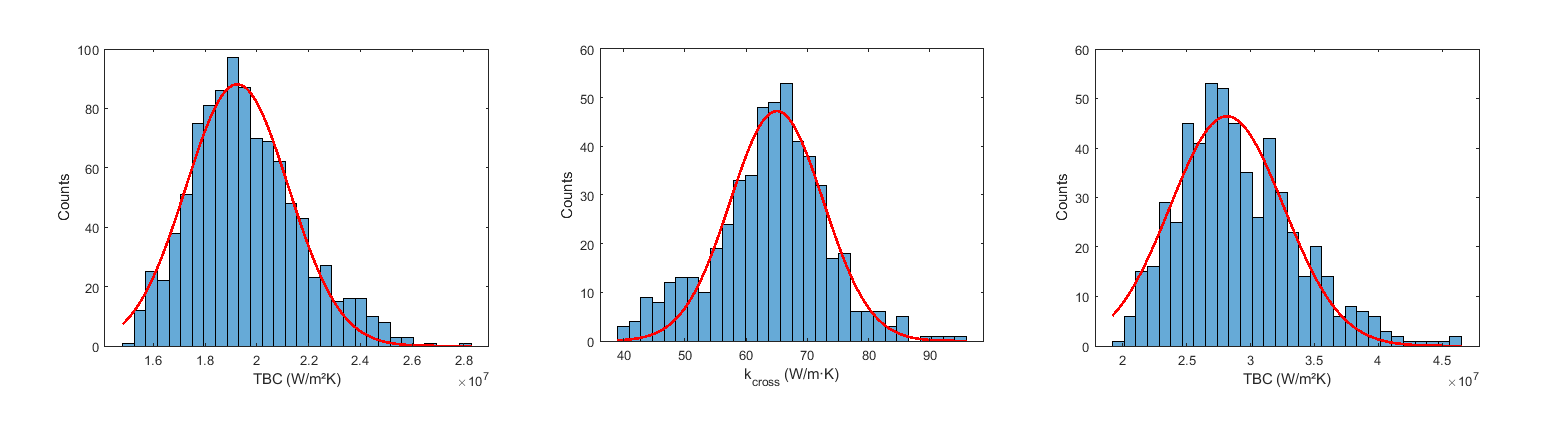}
    \caption{Plots of the Monte Carlo fitting algorithm with histogram counts representing the fitted values and error spread for the Au/GaN thermal boundary conductance (left), the cross-plane thermal conductivity of GaN (middle) fitted as $k_{\parallel} =$ (92.8 $\pm$ 22.0) W/m$\cdot$K, and the GaN/Si thermal boundary conductance (right) fitted as $G = (2.8 \pm 0.2)\times10^7$ W/m$^2\cdot$K.}
    \label{fig:S12}
\end{figure}

The error on the thermal conductivity and the TBC from FDTR quoted in the main text is representative of a confidence interval from performing the Monte Carlo fit 1000 times and obtaining the width of the spread in Fig. \ref{fig:S12}. As the experimental measurement is repeated on different spots on the sample, the standard deviation of the fitted spread value from different spots provides the true experimental error bar. In the text, we use the confidence interval as an upper bound to emphasize the error of the FDTR technique at a single spot.

\section{Literature on thermal conductivity of gallium nitride}

In this supplementary note, we provide a table of previous literature values of thermal conductivity of GaN in both bulk and thin film forms and on different substrates to serve as a comparison to our study. We note that the table is not comprehensive with the entire literature on GaN thermal transport. These values are plotted in Fig. 4c of the main text.

\begin{table}[h!]
\caption{Compilation of values (or ranges of values) of the thermal conductivity and thermal boundary conductance of GaN at room temperature on different substrates and thicknesses found in a subset of literature.}
\begin{tabular}{c|c|c|c|c}
\multicolumn{1}{l|}{\textbf{Substrate}} & \multicolumn{1}{l|}{\textbf{Thickness (nm)}} & \multicolumn{1}{l|}{\textbf{$k$ (W/m$\cdot$K)}} & \textbf{Reference} \\ \hline
 Si & 500 & 92.8 $\pm$ 22.0 & This work \\
 Bulk & Bulk & 240 $\pm$ 20 & \cite{jezowski2003,mion2006,simon2014,jezowski2015} \\
 Si & 1200 & 150 $\pm$ 22.5 & \cite{sarua2007} \\
 GaN & 250-2100 & 150-195 & \cite{koh2021} \\
 GaN/sapphire & 400-800 & 55-167 & \cite{koh2021} \\
 AlN/sapphire & 3190 & 162 $\pm$ 16 & \cite{li2020} \\
 Sapphire & 3520 & 190 $\pm$ 15 & \cite{li2020} \\ 
 4H-SiC & 14-940 & 2-118 & \cite{ziade2017} \\
 AlN/SiC & 900 & 167 $\pm$ 15 & \cite{cho2014} \\
 AlN/Si & 1300 & 185 $\pm$ 20 & \cite{cho2014}
\end{tabular}
\end{table}

\newpage
\bibliographystyle{apsrev4-1.bst}
\bibliography{references.bib}